# Spectrally flat and broadband double-pumped fiber optical parametric amplifiers

[1]J. M. Chavez Boggio, [1]J. D. Marconi, [2]S. R. Bickham, and [1]H. L. Fragnito

[1]*Optics and Photonic Research Center (Cepof), UNICAMP, Campinas, 13083-970, SP, Brazil*
[2]*Corning Incorporated, Science and Technology Division, One Science Center Drive, SP-TD-1, Corning, NY 14831, USA*
*jmchavez@ifi.unicamp.br*

**Abstract:** We study theoretically and experimentally spectrally flat and broadband double-pumped fiber-optical parametric amplifiers (2P–FOPAs). Closed formulas are derived for the gain ripple in 2P-FOPAs as a function of the pump wavelength separation and power, and the fiber non-linearity and fourth order dispersion coefficients. The impact of longitudinal random variations of the zero dispersion wavelength ($\lambda_0$) on the gain flatness is investigated. Our theoretical findings are substantiated with experiments using conventional dispersion shifted fibers and highly nonlinear fibers (HNLFs). By using a HNLF having a low variation of $\lambda_0$ we demonstrate high gain and flat spectrum (25 ± 1.5 dB) over 115 nm.

**OCIS codes:** (060.2320) Fiber optics amplifiers and oscillators; (060.2330) Fiber optics communications; (190.4370) Nonlinear optics, fibers.

## References and links

---

## 1. Introduction

High capacity dense wavelength division multiplexed (DWDM) systems require broadband optical amplifiers with low ripple gain spectrum. Raman amplifiers in a multi-wavelength pump configuration and Erbium doped fiber amplifiers (EDFAs) providing flat gain over ~100 nm have been demonstrated and are commercially available [1,2]. However, it has been predicted that in the near future, the required bandwidth would be of several hundreds of nanometers [3]. Furthermore, in future optical networks additional functionalities (for example, wavelength conversion for all-optical networking) instead of only amplification will be required. Therefore, there is an increased interest in devices with multifunctional capabilities that could operate over very broad bands with flat spectral response.

A fiber optical parametric amplifier (FOPA) has the potential of providing high gain over a very broad bandwidth and also offering other functionalities such as wavelength conversion, optical reshaping, wavelength exchange, phase conjugation, etc [4,5]. The single-pumped FOPA (1P-FOPA) can exhibit gain over large bandwidths but with rather poor uniformity [6-9]. This problem can be solved by concatenating fibers with different zero dispersion wavelengths ($\lambda_0$) and lengths [10-13], where numerical simulations predict gain bandwidths exceeding 200 nm with a ripple of 0.2 dB [13]. Double-pumped FOPAs (2P-FOPAs) can offer flat gain spectra in a single fiber [14-17]. Theoretically, it has been shown that a very flat gain spectrum can be obtained in fibers with positive fourth order dispersion coefficient ($\beta_4$) [15]. The flatness in a given bandwidth is improved by reducing the absolute value of $\beta_4$ and by increasing the fiber nonlinear coefficient ($\gamma$) and the pump power ($P_1 + P_2$) (or alternatively by reducing the fiber length) [15,17]. For example, uniform (ripple < 1 dB) gain over more than 300 nm is predicted in 2P-FOPAs employing fibers with $\beta_4 \sim 10^{-6}$ ps$^4$/km and $\gamma = 15$ W$^{-1}$/km [18]. Experimental results, however, have proved that theoretical predictions are too optimistic in general. Using a highly nonlinear dispersion shifted fiber (HNLDSF) with positive $\beta_4$, a gain ripple of 3 dB over 33.8 nm bandwidth has been demonstrated [19] (and 41.5 nm for ASE amplification [20]). Recently 47 nm of flat gain was demonstrated using conventional DSFs and 71 nm using a HNLF having negative $\beta_4$ [21,22].

Several factors conspire against obtaining flat-broadband FOPAs in practice. One is the lack of highly nonlinear fibers with the desired dispersion coefficients (for example, the lowest $\beta_4$ reported is ~$3\times10^{-5}$ [23], while the highest $\gamma$ is 30 W$^{-1}$/km [24]). Another factor is the fact that a real fiber exhibits random variations of the zero-dispersion wavelength along its length [16, 25-27]. Calculations reported in [26] predicted that the bandwidth of flat operation of 2P-FOPAs would be limited to less than 100 nm due to unavoidable fluctuations of $\lambda_0$. Still another factor is the effect of polarization mode dispersion (PMD) [28-31], which tends to produce distortions in the gain spectrum when the PMD parameter of the fiber and the pump separation are large [29].

In this paper we study theoretically and experimentally spectrally flat and broadband 2P–FOPAs. The main purpose of the theoretical part of this paper is to obtain analytical expressions of the gain ripple for the various types of 2P-FOPA gain spectra, which we classify by their number of extrema in sections 2, 3, and 4. In section 5 we analyze the impact of longitudinal variations of $\lambda_0$ on gain flatness. In sections 6 and 7 we present our experimental results. By using a well designed highly nonlinear fiber having a variation of $\lambda_0$ of ~0.1 nm we demonstrate high gain and flat spectrum (25 ± 1.5 dB) over 115 nm. Finally, in section 8 we draw our conclusions.

## 2. The extrema of the 2P-FOPA gain spectrum and calculation of the gain ripple

The FWM process responsible for parametric gain in a 2P-FOPA satisfies $\omega_1 + \omega_2 = \omega_s + \omega_i$; where $\omega_1$, $\omega_2$, $\omega_s$, and $\omega_i$ are the pumps, signal and idler frequencies, respectively. The propagation constant mismatch of this FWM process is given by

$$\Delta\beta = \beta_{2c}[\Delta\omega_s^2 - \Delta\omega_p^2] + \beta_{4c}[\Delta\omega_s^4 - \Delta\omega_p^4]/12 + \ldots \quad (1)$$

where $\omega_c = (\omega_1 + \omega_2)/2$, $\Delta\omega_s = \omega_s - \omega_c$, $\Delta\omega_p = \omega_1 - \omega_c$, and $\beta_{2c} = \beta_2(\omega_c)$ and $\beta_{4c} = \beta_4(\omega_c)$ are the second and fourth order dispersion coefficients evaluated at $\omega_c$, respectively. The pumps provide a nonlinear contribution to the phase of the waves, so that the total propagation constant mismatch is $\kappa = \Delta\beta + \gamma(P_1 + P_2)$, where $P_1$ and $P_2$ are the pump powers, and $\gamma$ is the fiber nonlinear coefficient. The scope of this paper is restricted to fibers with conventional dispersion profiles, having only one $\lambda_0$ and quartic dispersion relation in the spectral region of interest (i.e., we neglect fifth and higher order dispersion terms, then $\beta_{4c} = \beta_4$ is frequency independent). If the fiber loss can be neglected, the parametric gain, $G$, is given by [17]

$$G = 1 + \left(\frac{x_0 \sinh x}{x}\right)^2, \quad (2a)$$

$$x = x_0 \sqrt{1 - \left(\frac{\kappa}{2\gamma P_0}\right)^2}, \quad (2b)$$

where $x_0 = \gamma P_0 L$, $L$ is the fiber length, and $P_0 = 2\sqrt{P_1 P_2}$ .

*2.1 The extrema of the 2P-FOPA gain spectrum*

As a first step to analyze the gain flatness of 2P-FOPAs, we calculate the extrema of $G(\omega_s)$ that are obtained from the zeros of the derivative of $G$ with respect to $\Delta\omega_s$

$$\frac{\partial G}{\partial \Delta\omega_s} = -\frac{1}{2} x_0^2 L^2 f(x) \frac{\partial \Delta\beta}{\partial \Delta\omega_s} \kappa = 0 \quad (3a)$$

$$f(x) = \sinh(x)\left(x\cosh(x) - \sinh(x)\right)/x^4 . \quad (3b)$$

The gain is exponential when $x$ is real and in this case we have that $f(x)$ ($\geq f(0) = 1/3$) is monotonic crescent. The extrema are then given by $\partial\Delta\beta/\partial\Delta\omega_s = \Delta\omega_s\left(2\beta_{2c} + \beta_4 \Delta\omega_s^2/3\right) = 0$ and $\kappa = 0$. The zeros of $\partial\Delta\beta/\partial\Delta\omega_s$ are located at $\Delta\omega_s = 0$ and at $\Delta\omega_s = \pm\sqrt{-6\beta_{2c}/\beta_4}$ , while the zeros of $\kappa$ are at

$$\Delta\omega_s = \pm\Delta\omega_p \sqrt{\frac{-6\beta_{2c}}{\beta_4 \Delta\omega_p^2} \pm \sqrt{\left(\frac{6\beta_{2c}}{\beta_4 \Delta\omega_p^2} + 1\right)^2 - \frac{12\gamma(P_1 + P_2)}{\beta_4 \Delta\omega_p^4}}} . \quad (4)$$

In principle, there could be four roots of $\kappa = 0$. To know if the extrema are maxima or minima (absolute or local) we calculate the second derivative of $G$

$$\frac{\partial^2 G}{\partial \Delta\omega_s^2} = -\frac{1}{2} x_0^2 L^2 \left\{ \kappa \frac{df}{dx} \frac{\partial x}{\partial \Delta\omega_s} \frac{\partial \Delta\beta}{\partial \Delta\omega_s} + f(x)\left[ \left(\frac{\partial \Delta\beta}{\partial \Delta\omega_s}\right)^2 + \kappa \frac{\partial^2 \Delta\beta}{\partial \Delta\omega_s^2} \right] \right\} \quad (5)$$

From Eq. 5 we can see that the zeros of $\kappa$ are all absolute maxima. These are points of perfect phase matching where we have $G = 1 + \sinh^2 x_0$. The zeros of $\frac{\partial \Delta\beta}{\partial \Delta\omega_s}$, are local maxima if $\kappa \frac{\partial^2 \Delta\beta}{\partial \Delta\omega_s^2} = \kappa \left\{ \beta_{2c} + \frac{\beta_4}{2} \Delta\omega_s^2 \right\} > 0$.

Thus there can be spectra having 7 extrema (four maxima and three minima), 5 (three maxima and two minima), 3 (two maxima and one minimum), or 1 (one maximum). The extremum at $\Delta\omega_s = 0$ always exists, while the existence of the other extrema will depend on the particular values of the FOPA parameters $\beta_{2c}$, $\beta_4$, $\gamma P_0$, and $\Delta\omega_p$. (For example, it is easy to show that a necessary condition for the existence of the extrema at $\Delta\omega_s = \pm\sqrt{-6\beta_{2c}/\beta_4}$ is that $\beta_{2c}$ and $\beta_4$ have opposite signs).

It would be useful for FOPA design to have expressions of the gain ripple for these kinds of spectra. As noticed in [15] $\kappa$ as a function of $\Delta\omega_s$, being a fourth order polynomial, has minimum ripple in a given region ($|\Delta\omega_s| < \Delta\omega_t$) if it is proportional to the Chebyshev polynomial $T_4 = 1 - 8(\Delta\omega_s/\Delta\omega_t)^2 + 8(\Delta\omega_s/\Delta\omega_t)^4$. This approach is very useful for fibers with $\beta_4 > 0$ and is further analyzed in section 3.1. If $\beta_4 < 0$, it follows from Eq. 4 that the two outermost roots of $\kappa = 0$ always exist and are located outside the pumps ($|\Delta\omega_s| > |\Delta\omega_p|$). The Chebyshev bandwidth, $\Delta\omega_t$, is then larger than $\Delta\omega_p$, i.e. includes always the pump frequencies. In practice, however, as shown in the experimental part, the region around the pumps cannot be used in general for parametric amplification, since other 'spurious' nonlinear effects are very strong in those regions. Around the pumps, the combined actions of processes satisfying $\omega' = 2\omega_1 - \omega_s$ and $\omega' = \omega_1 - \omega_2 + \omega_i$ drastically perturb the 2P-FOPA, generally reducing the gain [17]. Furthermore, as shown in appendix A, these are regions of strong crosstalk when the 2P-FOPA is used for DWDM applications. In order to avoid these 'spurious' effects one has to limit the operation of the 2P-FOPA to a spectral region smaller than $\Delta\omega_p$, say $\Delta\omega_s < b\Delta\omega_p$ ($0 < b < 1$). Minimizing the gain ripple in this reduced region cannot be treated with the fourth order Chebyshev polynomial approach. This is considered next.

### *2.2 Gain ripple in 2P-FOPAs*

The procedure for the calculation of the gain ripple in the various kinds of spectra is introduced in this subsection with an example of a fiber with $\beta_4 < 0$. Figure 1 shows a set of gain spectra obtained by tuning $\lambda_c$ from 1544.56 to 1544.87 nm in steps of 0.039 nm. We considered a FOPA with $L = 420$ m, $\gamma = 15$ (W-km)$^{-1}$, $P_1 = P_2 = 0.25$ W, $\lambda_2 - \lambda_1 = 100$ nm, third order dispersion $\beta_{3c} = \beta_3(\omega_c) = 0.065$ ps$^3$/km, $\beta_4 = -8.5\times10^{-5}$ ps$^4$/km, $\lambda_2 = 1595$ nm, and $\lambda_0 = 1545$ nm. The shortest set of $\lambda_c$ values result in spectra with 7 extrema (showed in Fig. 1a), then increasing $\lambda_c$ yields spectra with 5 extrema (Fig. 1b). A further increase in $\lambda_c$ results in spectra with 3 extrema, but with very low gain for these FOPA parameters. (Spectra with only one extremum at $\Delta\omega_s = 0$, as can be straightforwardly derived from Eq. (4), only exist in fibers with $\beta_4 > 0$.)

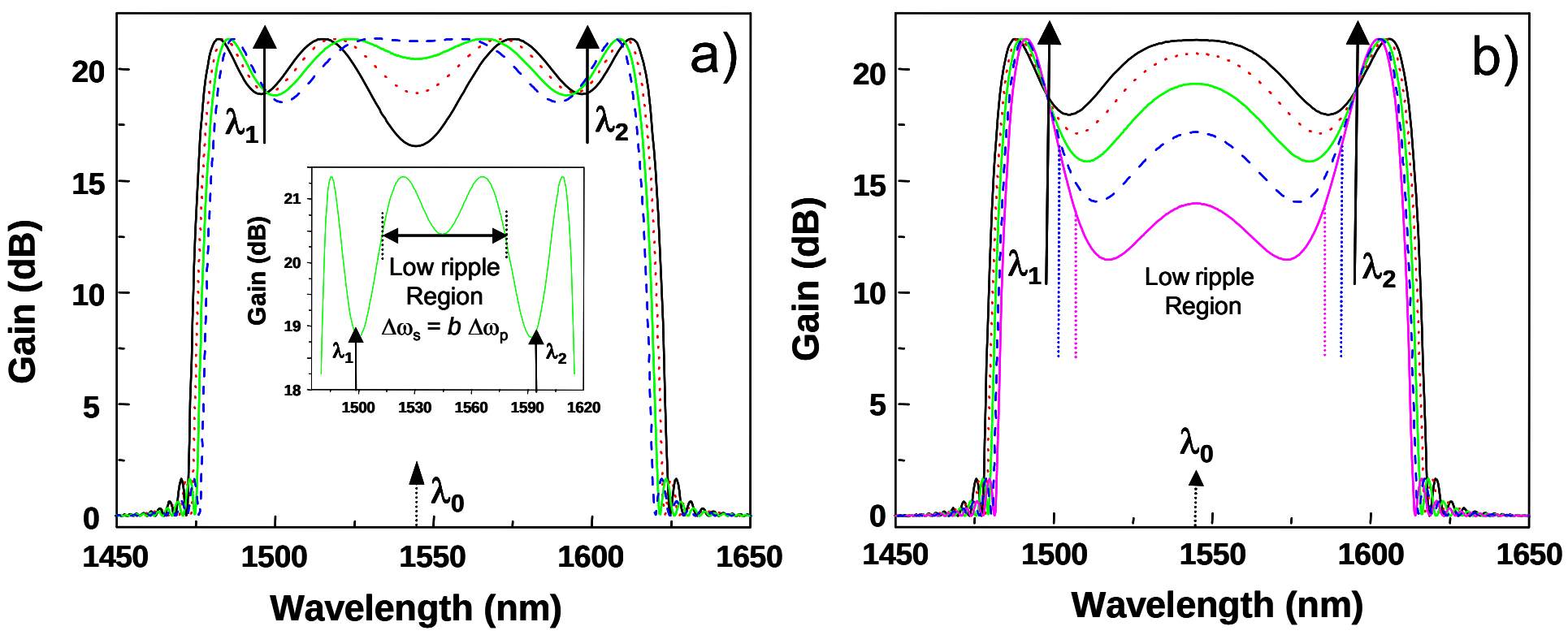

Fig. 1. Gain spectra calculated using Eq. (2). By tuning $\lambda_c$ from shorter to longer wavelengths we have the spectra in (a) Black ($\lambda_c$ = 1544.56 nm), red ($\lambda_c$ = 1544.6 nm), green ($\lambda_c$ = 1544.64 nm) and blue ($\lambda_c$ = 1544.68 nm). (b) Black ($\lambda_c$ = 1544.72 nm), red ($\lambda_c$ = 1544.75 nm), green ($\lambda_c$ = 1544.79 nm), blue ($\lambda_c$ = 1544.83 nm), and magenta ($\lambda_c$ = 1544.87 nm).

The spectral region near the pumps should be avoided due to cross-talk (see Appendix A). Note that the spectra depicted in green in Fig. 1(a) exhibit low ripple over a bandwidth, which not includes the region near the pumps. In comparison, over comparable bandwidths, the spectra in black and blue exhibit poorer gain flatness. In the case of Fig. 1(b), the spectra in blue and magenta exhibit also regions of lower ripple (but with a decreased gain if compared with the 7 extrema case) if compared with spectrum in blue.

To calculate the gain ripple of the low ripple regions observed in Fig. 1, we need to know the value of $\beta_{2c}$ for each case. This is obtained by equalizing the gain at $\Delta\omega_s = 0$, which is a minimum (maximum) in spectra with seven (five) extrema, with the gain at a frequency $\Delta\omega_s = b\Delta\omega_p$, where $0 < b \leq 1$. From Eq. 2 we note that this corresponds to equalize $\kappa^2$ at these wavelengths, i.e. $\kappa\,(\Delta\omega_s = 0) = \pm\kappa\,(\Delta\omega_s = b\Delta\omega_p)$. This procedure yields two possible values of $\beta_{2c}$:

$$\beta_{2c} = -\beta_4 \Delta\omega_p^2 b^2 / 12 \tag{6a}$$

$$\beta_{2c} = -\frac{\beta_4 \Delta\omega_p^2}{12\left(2-b^2\right)}\left[\frac{1}{\xi} - (2-b^4)\right] \tag{6b}$$

where

$$\xi = \frac{\beta_4 \Delta\omega_p^4}{24\gamma(P_1+P_2)}\,. \tag{7}$$

By substituting each value of $\beta_{2c}$ in Eq. (4) we note that with the value in Eq. 6(a) we can have only two roots of $\kappa = 0$ (i.e. spectra with five extrema), while with the value of $\beta_{2c}$ in 6(b) we have four roots of $\kappa = 0$ (spectra with seven extrema). Thus, with those values of $\beta_{2c}$ it is possible to calculate $\kappa$ (and the gain) at the extrema. For example, with the $\beta_{2c}$ in Eq. 6(a) we can calculate the gain at $\Delta\omega_s = 0$ (which is the maximum, $G_{max}$) and also at $\Delta\omega_s = \pm\sqrt{-6\beta_{2c}/\beta_4}$ (which is the minimum, $G_{min}$). From these values we obtain the gain ripple: $\Delta G = G_{max} - G_{min}$. In the same way with $\beta_{2c}$ in Eq. 6(b) we can calculate $G_{min}$ at $\Delta\omega_s$ = 0, and knowing that $G_{max} = 1 + \sinh^2 x_0$ in the spectra with 7 extrema, we can then find $\Delta G$.

It is convenient, in order to have tractable expressions of $G_{max}$ and $G_{min}$, to take the limiting case $\sinh^2 x \sim e^{2x}/4$, with error < 3 % for $x > 2$. Using this approximation, $G$ in decibel units is

$$G_{dB} \cong 8.7x_0\sqrt{1-\left(\frac{\kappa}{2\gamma P_0}\right)^2} - 10\log(1-\left(\frac{\kappa}{2\gamma P_0}\right)^2) - 6 \tag{8}$$

In sections 3 and 4 we analyze $\Delta G$ for the most representative types of gain spectra.

## 3. Gain ripple in 2P-FOPA spectra with seven extrema

### *3.1 The fourth order polynomial Chebyshev gain spectrum*

In this subsection we calculate the gain ripple of the Chebyshev spectrum as the parameters $\beta_4$, $\gamma(P_1 + P_2)$, and $\Delta\omega_p$ are varied. This spectrum occurs when the three local minima have the same gain, i.e. when $\kappa^2$ is the same when evaluated at $\Delta\omega_s = 0$ or at $\Delta\omega_s = \pm\sqrt{-6\beta_{2c}/\beta_4}$. The condition $\kappa(\Delta\omega_s = 0) = -\kappa(\Delta\omega_s = \pm\sqrt{-6\beta_{2c}/\beta_4})$ results in the Chebyshev spectrum characterized by

$$\beta_{2c} = \frac{-\beta_4\Delta\omega_p^2}{3}\left[1-\sqrt{\frac{1}{2}+\frac{1}{4\xi}}\right] \tag{9}$$

Figure 2(a) shows the gain spectra obtained with this value of $\beta_{2c}$ for fibers with $\beta_4 > 0$ (blue line) and $\beta_4 < 0$ (black line) for a 2P-FOPA with the same parameters used in Fig. 1 except that now $\beta_4 = \pm 8 \times 10^{-5}$ ps$^4$/km. These parameters result in $x_0 = 3.15$ and $\xi = \pm 1.07$. The fiber with $\beta_4 > 0$ exhibits a ripple of 0.045 dB over a region, given by $\Delta\omega_t = (-12\beta_{2c}/\beta_4)^{1/2}$, which we call the Chebyshev bandwidth and is indicated by dotted blue lines. The fiber with $\beta_4 < 0$ gives a much larger ripple of 3.6 dB.

With the value of $\beta_{2c}$ from Eq. (9) the phase mismatch at minimum gain is $\kappa_{min}/2\gamma P_0 = (\sqrt{2|u|} - \sqrt{0.5\,\mathrm{sgn}(\xi)+|\xi|})^2$. The sign function of $\xi$, $sgn(\xi)$, is negative (positive) in fibers with $\beta_4 < 0$ (> 0). We then substitute this value of $\kappa_{min}/2\gamma P_0$ in Eq. 8 to obtain $G_{min}$. Since the maximum gain (at $\kappa = 0$) is given in dB by $G_{max} \cong 8.7x_0 - 6$, the gain ripple is

$$\Delta G_{dB} \cong 8.7x_0\left\{1-\sqrt{1-\left(\sqrt{2|\xi|}-\sqrt{0.5\,\mathrm{sgn}(\xi)+|\xi|}\right)^4}\right\} + 10\log\left(1-\left(\sqrt{2|\xi|}-\sqrt{0.5\,\mathrm{sgn}(\xi)+|\xi|}\right)^4\right) \tag{10}$$

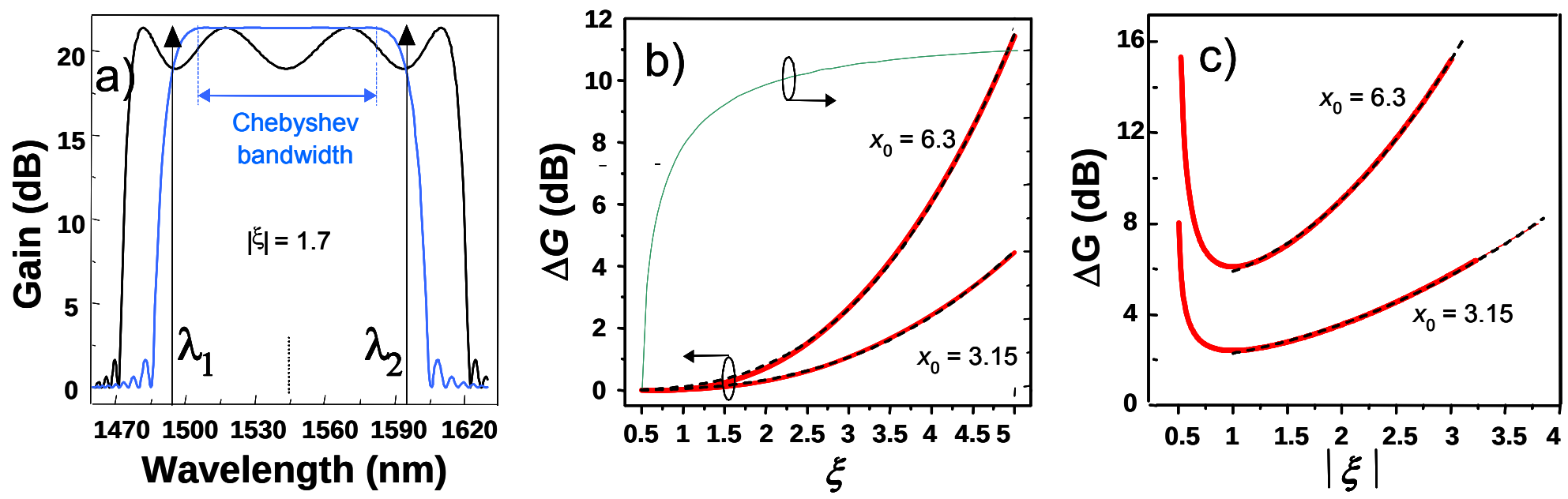

Fig. 2. (a) Gain spectrum with Chebyshev shape for positive (blue line) and negative (black line) $\beta_4$. (b) $\Delta G_{dB}$ as a function of $\xi$ for $\beta_4 > 0$ and (c) $\Delta G_{dB}$ as a function of $\xi$ for $\beta_4 < 0$. Solid red lines: $\Delta G$ as calculated with Eq. (10) for $G_{max}$ = 21.35 dB and $G_{max}$ = 48 dB. Dotted black lines: fittings with the expressions quoted in table I.

Equation (10) expresses the gain ripple as a function of the parameter $\xi$. Before discussing the results from Eq. (10), it is important to mention the range of FOPA parameters for which the preceded analysis is consistent and meaningful. The existence of four roots in $\kappa$ occurs only if $\beta_{2c}$ and $\beta_4$ have opposite signs. From Eq. (9) it is straightforward to see that this occurs only when $|\xi| \geq ½$. Therefore, Eq. (10) is not valid for $|\xi| < ½$.

Figures 2(b) and 2(c) show $\Delta G$ calculated in fibers with positive and negative $\beta_4$ values, respectively and for two representative values of the parametric gain: $G_{max}$ = 21.35 dB ($x_0$ = 3.15) and $G_{max}$ = 48 dB ($x_0$ = 6.3). In general, decreasing $\xi$ flattens the 2P-FOPA and the Chebyshev spectrum can offer very low ripple when $\beta_4 > 0$. As a specific example, we consider a FOPA characterized by: $\gamma = 30$ (W-km)$^{-1}$, $P_1 + P_2 = 0.6$ W, $\beta_4 = 1\times10^{-5}$ ps$^4$/km, pump separation of ~ 33 THz (250 nm centered at 1535 nm). These values results in $\xi = 2.67$, i.e. $\Delta G$ ~ 0.8 dB. This small ripple corresponds to a flat gain spectrum over nearly 250 nm. In Fig. 2(b) we have also plotted the Chebyshev bandwidth normalized to the pump separation as a function of $\xi$. In this case a bandwidth larger than 0.85 is obtained if $\xi > 1.5$.

For fibers having $\beta_4 < 0$, the smallest ripple (2.4 dB for $G_{max}$ = 21.35 dB and 6 dB for $G_{max}$ = 48 dB) is obtained when $|\xi| = 1$. For $½ < |\xi| < 1$, the ripple increases.

Table I. Expression for fitting $\Delta G$.

| | $\beta_4 < 0$ | $\beta_4 > 0$ |
|---|---|---|
| $x_0 = 3.15$ | $\Delta G = 1.9 + 0.42\xi^2$ | $\Delta G = 0.05\xi^{2.85}$ |
| $x_0 = 6.3$ | $\Delta G = 5.1 + 0.8\xi^{2.3}$ | $\Delta G = 0.1\xi^{2.9}$ |

Even though Eq. (10) expresses the gain ripple as a complicated function of $\xi$, it is possible to approximate $\Delta G$ with simple expressions of the type $\Delta G_{dB} = a \times \xi^p$ (or $\Delta G_{dB} = a_0 + a \times \xi^p$), where $a_0$, $a$, and $p$ are constants. Examples of these power law fits are represented by dotted lines in Figs. 2(b) and 2(c). For the case $\beta_4 < 0$ the fit was for $|\xi| > 1$. Table I quotes the respective values of $a_0$, $a$, and $p$. These simple expressions can be used as a rule of thumb to estimate the amount of increase (or decrease) in $\Delta G$ by increasing (or decreasing) $\xi$. For example, when $G_{max}$ = 48 dB and $\beta_4 > 0$, increasing $\xi$ by a factor of 2 (for instance by increasing $\beta_4$ by a factor of 2), should lead to a factor of $2^{2.9}$ ~ 8 increase in $\Delta G$.

*3.2 Gain spectrum with seven extrema and arbitrary shape*

The gain spectrum with Chebyshev shape in fibers with $\beta_4 < 0$ had a rather poor flatness, but the gain ripple can be minimized for the other spectral shapes discussed in Fig. 1 (a). Figure 3(a) shows the gain spectrum obtained with the same parameters as in Fig. 2(a) when the region of minimization is $b = \Delta\omega_s/\Delta\omega_p = 0.85$.

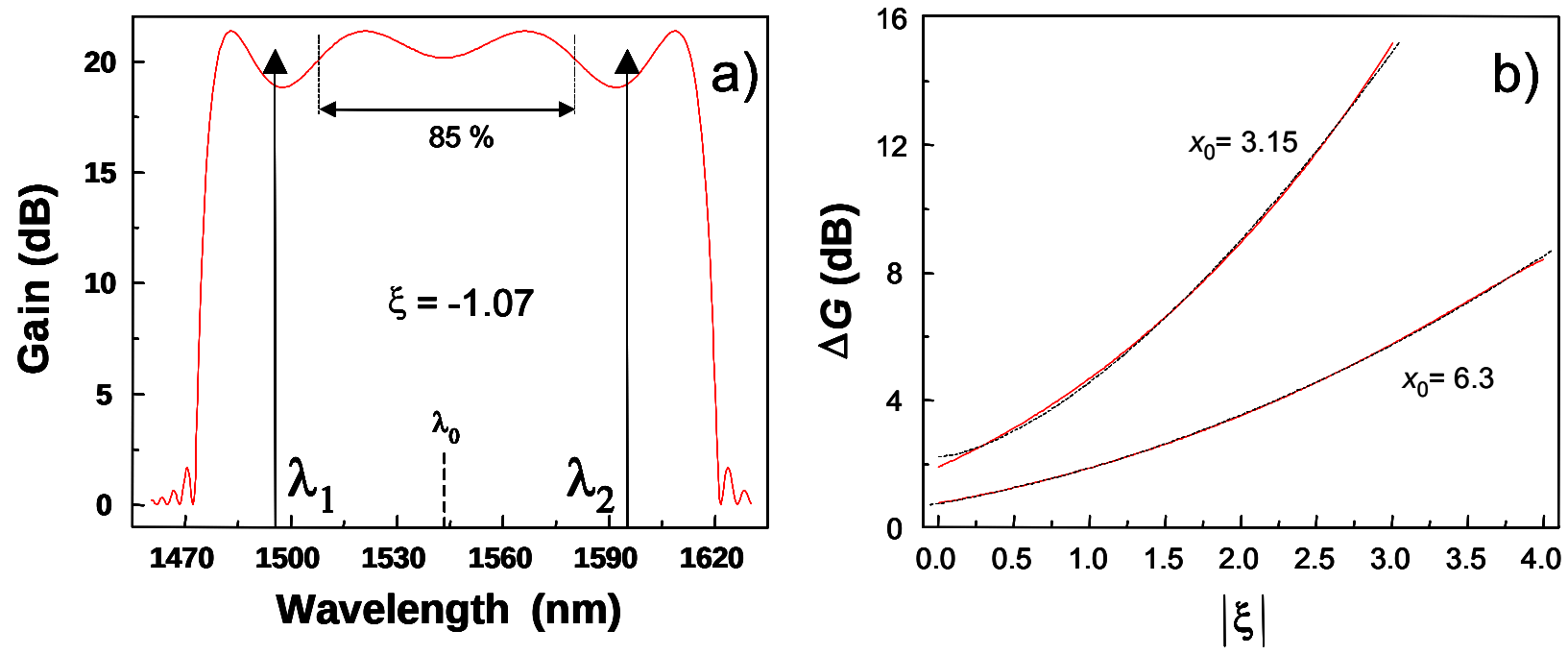

Fig. 3. (a) Gain spectrum when the ripple is minimized in a region $\Delta\omega_s = 0.85\Delta\omega_p$. (b) In red lines: Calculated $\Delta G_{dB}$ as a function of $\xi$ for $b = 0.85$ for two values of $x_0$. Black dotted lines: Power law fits to $\Delta G$. (For $x_0 = 3.15$ we have $\Delta G = 0.9 + 0.9|\xi|^{1.53}$, while for $x_0 = 6.3$ we have $\Delta G = 2.2 + 2.3|\xi|^{1.5}$.)

The phase mismatch at minimum gain can be calculated using the value of $\beta_{2c}$ in Eq. 6(b) as a function of the region of ripple minimization, $b$:

$$\frac{\kappa_{mín}}{2\gamma P_0} = 0.5\left(\frac{b^2}{b^2-2}\right)\left(0.5-\xi\left(1-b^2\right)\right) \qquad (11)$$

To calculate the gain ripple we note that the maximum gain is $G_{\max} \cong 8.7x_0 - 6$, while the minimum gain is calculated by combining Eqs. (11) and (8). In Figure 3(b) we plot the gain ripple for the case $b = 0.85$ and for two values of $x_0 = 3.15$ and $x_0 = 6.3$. Comparing these results to those shown in Fig. 2, for values of $|\xi| > 1.5$ (where the gain ripple is high) the two results are very similar for both $x_0 = 3.15$ and 6.3. For values $|\xi| < 1.5$ the spectrum analyzed in this subsection exhibits a smaller ripple. This means that it is possible to reduce the ripple by slightly reducing the bandwidth of amplification. (Note in Eq. 11 that the $\kappa$ is reduced as long as we reduce $b$.)

## 4. Gain ripple in 2P-FOPA spectra with five extrema

In this section we study minimization of the gain ripple in spectra having five extrema in fibers with $\beta_4 < 0$ (The case of fibers with $\beta_4 > 0$ is discussed in Appendix B). Figure 4 shows two typical gain spectra with identical FOPA parameters ($x_0$, $\beta_4$, and $\Delta\omega_p$) as in Figs. 2(a) and 3(a). The spectra were obtained for two different ways of minimizing the gain ripple: the solid line corresponds to equalizing the gain at $\Delta\omega_s = 0$ with that at $\Delta\omega_s = \Delta\omega_p$, while the dashed line is obtained by equalizing the gain at $\Delta\omega_s = 0$ with that at $\Delta\omega_s = 0.85\Delta\omega_p$. This equalization leads to the value of $\beta_{2c}$ in Eq. 6(a) from which it is possible to calculate the phase mismatch at $\Delta\omega_s = 0$ (maximum) and at $\Delta\omega_s = \pm\sqrt{-6\beta_{2c}/\beta_4}$ (minima):

$$\frac{\kappa_{max}}{2\gamma P_0} = \xi(b^2 - 1) + \frac{1}{2}, \tag{12a}$$

$$\frac{\kappa_{min}}{2\gamma P_0} = \xi b^2 (1 - \frac{b^2}{4}) - \xi + \frac{1}{2}. \tag{12b}$$

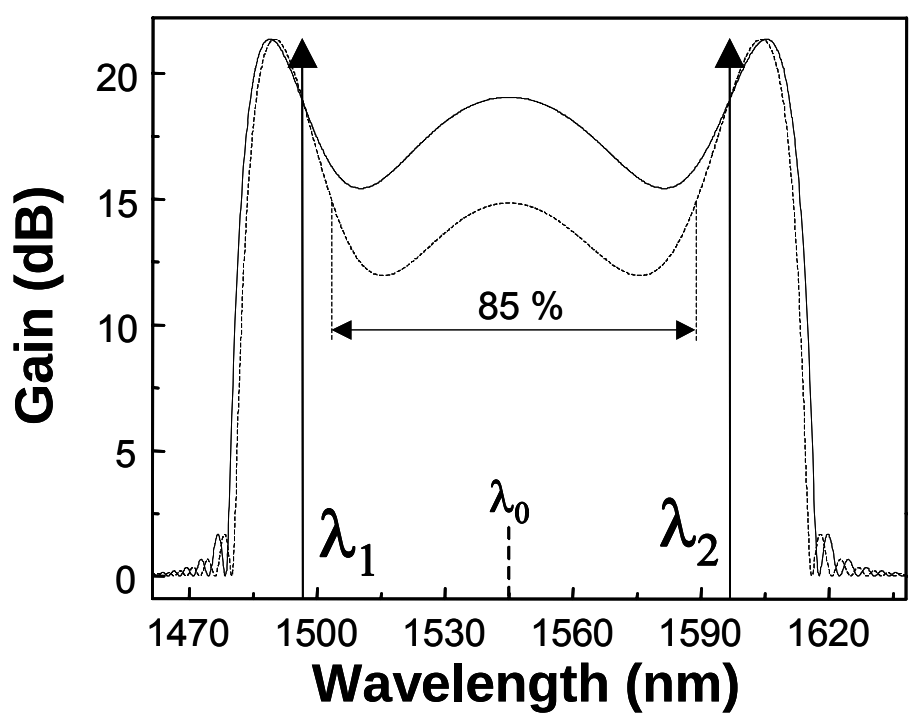

Fig. 4. Gain spectra having 5 extrema. The gain ripple was minimized in the region $\Delta\omega_s = \Delta\omega_p$ (solid line) and $\Delta\omega_s = 0.85\Delta\omega_p$ (dashed line).

Equations 12(a) and 12(b) were used to calculate $G_{max}$, $G_{min}$, and then $\Delta G$ as a function of $b$. Figures 5(a) and 5(b) show $\Delta G$ for $b$ = 1 and 0.85, respectively. As in the case of spectra with 7 extrema, it is apparent that low ripple is obtained for small values of $|\xi|$. Also, the gain ripple is slightly smaller for a smaller value of $b$. However, this slight improvement in flatness is obtained by reducing the overall gain as can be observed in Fig. 4. It is interesting comparing Figs. 5(b) and 3(b) (i.e. when the ripple is minimized in the region $\Delta\omega_s = 0.85\Delta\omega_p$): if $|\xi| < 0.5$ the case of spectra with 5 extrema produces a flatter spectrum if compared with the seven extrema case; on the other hand, if $|\xi| > 0.5$ similar values of $\Delta G$ for both cases are obtained when $x_0 = 3.15$; finally, when $x_0 = 6.3$ the spectrum with 7 extrema exhibit a flatter gain.

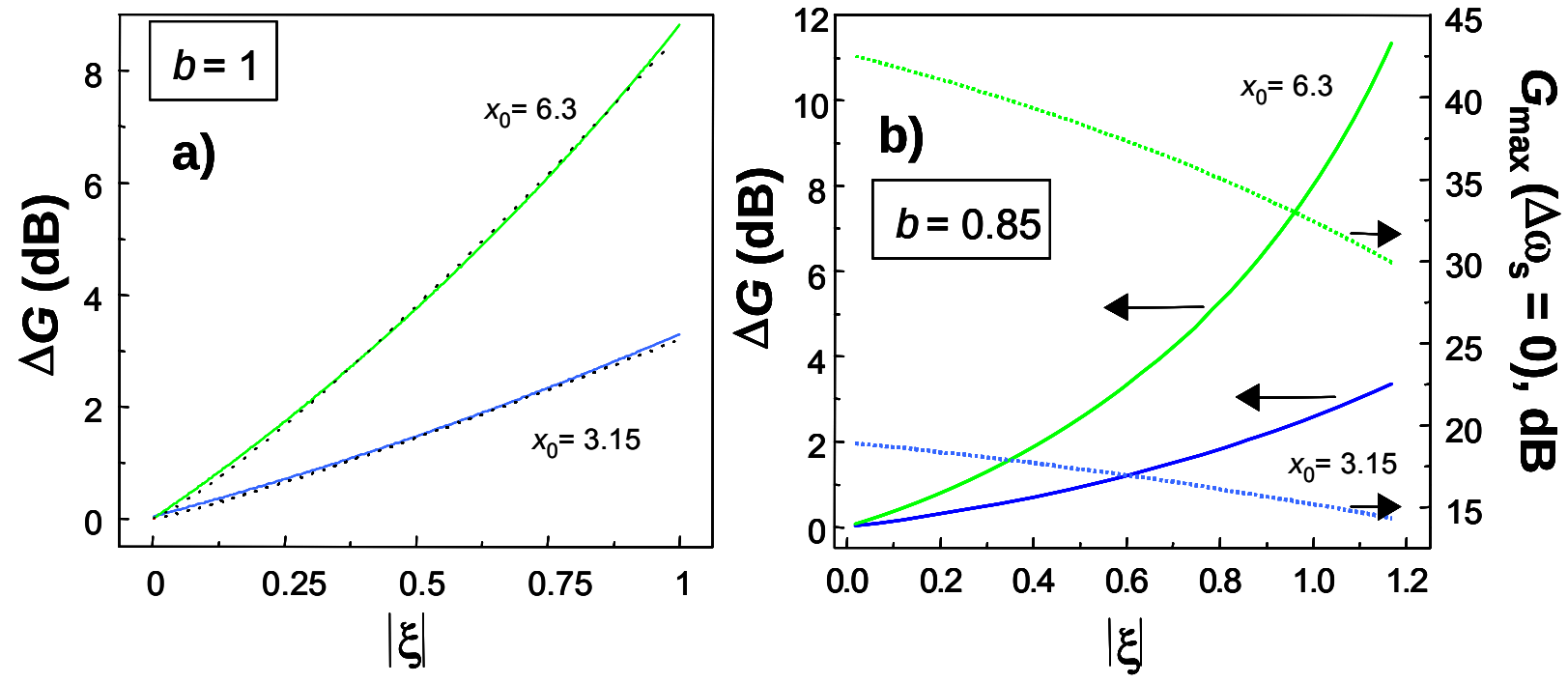

Fig. 5. $\Delta G_{dB}$ as a function of $\xi$ when the ripple is minimized in the region (a) $\Delta\omega_s = \Delta\omega_p$. (b) $\Delta\omega_s = 0.85\Delta\omega_p$. The dotted lines in Figure (a) show the power law fits to $\Delta G$. For $x_0 = 3.15$ we have $\Delta G = 3.2\ |\xi|^{1.15}$, while for $x_0 = 6.3$ we have $\Delta G = 8.7\,|\xi|^{1.2}$. The value of $G_{max}$ is 19 dB for the case $b = 1$, while for the case $b = 0.85$ depends on $\xi$ and we have plotted in Fig. (b).

Note in Fig. 5(a) that $\Delta G$ scales nearly linearly with $\xi$. It can be shown that the exact expression of $\Delta G$ for the optimised spectrum with 5 extrema and $b = 1$ is

$$\Delta G_{dB} \cong 4.3\sqrt{3}x_0\left(1-\sqrt{1+\frac{\xi}{3}-\frac{\xi^2}{12}}\right)+10\log\left(1+\frac{\xi}{3}-\frac{\xi^2}{12}\right) \tag{13}$$

For $|\xi| \ll 1$ Eq. (13) reduces to $\Delta G_{dB} \approx (-1.25x_0 + 1.45)\xi$. In very high gain amplifiers ($x_0 \gg 1$) the ripple becomes independent of pump power: for example, if $P_1 = P_2$, the limiting ripple is $\Delta G_{\mathrm{dB}} \approx -0.05\beta_4 \Delta\omega_p^4 L$.

## 5. Influence of variations of $\lambda_0$ and polarization mode dispersion

In general, there may be small random fluctuations of core radius and refractive index along the fiber, resulting in fluctuations of $\lambda_0$ that influence the efficiency of parametric amplifiers. In order to study the effects of variations of $\lambda_0$, we numerically solved the signal propagation Eqs. given in Ref. [26] by dividing the fiber in 5000 segments of length $\Delta z$. In each segment of fiber we defined a variation of the zero dispersion wavelength as $\lambda_0(z_k) = \langle\lambda_0\rangle + \delta_{\lambda 0}(z_k)$, where $k = 1, 2,.., 5000$, and the random variation $\delta\lambda_0(z_k)$ was generated using [33]

$$\delta\lambda_0(z_k) = \exp(-\Delta z / L_c)\delta\lambda_0(z_{k-1}) + \sigma_{\lambda 0}\sqrt{1-\exp(-2\Delta z / L_c)} \times r(k)\,, \tag{14}$$

where $\Delta z = z_k - z_{k-1}$, $L_c$ is a parameter related to the correlation length of the random process, and $r_k$ is a computer generated random number with normal distribution (zero mean and unit variance). By using this definition, $\delta\lambda_0(z_k)$ is a Gaussian stochastic process with expected values of $\langle\delta\lambda_0\rangle = 0$, correlation length $L_{\mathrm{corr}} = L_c(1 - e^{-L/Lc})$, and standard deviation $\sigma_{\lambda 0}$.

The gain ripple was calculated as a function of the standard deviation of the variation of $\lambda_0$ as follows: Eq. (14) was first used to generate a set of 25 to 35 simulated fibers for each value of $\sigma_{\lambda 0}$. In order to obtain the minimum ripple in each fiber, the gain spectrum was calculated for 60 pump locations by finely tuning of one of the pumps in a range of 1.2 nm and then keeping the flattest gain spectrum. Note that a similar procedure is employed in laboratory experiments to minimize the ripple. We then obtained the $\Delta G$ for each fiber and calculated the average of those 25-35 $\Delta G$ values.

### *5.1 The influence of third order dispersion on the impact of $\lambda_0$ fluctuations*

The 2P-FOPA parameters in our numerical simulations are $\gamma(P_1 + P_2) = 28$ km$^{-1}$, $L = 0.2$ km, $\beta_4 = -2 \times 10^{-4}$ ps$^4$/km, and average zero dispersion wavelength $\langle\lambda_0\rangle = 1570$ nm. The pumps are located at $\lambda_1 \cong 1520$ nm and $\lambda_2 \cong 1621$ nm, so the wavelength separation is ~100 nm. We assumed $L_c = 100$ m, then the correlation length is $L_{corr} \cong 86.5$ m. With this set of parameters $\xi \cong -0.6$ and we considered a gain spectrum of the type having 5 extrema. We did simulations for two values of the third order dispersion. Figure 6(a) shows the gain ripple as a function of $\sigma_{\lambda 0}$ for $\beta_3(\omega_0) = \beta_{30} = 0.065$ ps$^3$/km (red squares) and $\beta_{30} = 0.0325$ ps$^3$/km (black squares). Several interesting features can be observed. For both values of $\beta_{30}$, the ripple decreases as the variation of $\lambda_0$ increases reaching a minimum value before increasing strongly. This means that for this kind of spectrum, adequate amounts of variations of $\lambda_0$ tend to flatten the gain spectrum (the ripple was reduced from 4.3 dB to ~1.6 dB).

A second interesting feature is that the impact of the variation of $\lambda_0$ depends on the value of $\beta_{30}$: reducing $\beta_3$ by a factor of two allows $\sigma_{\lambda 0}$ to increase by a factor of two in order to have the same impact on gain ripple. Figure 6(b) shows a typical example of the 25 realizations (25 simulated fibers) having $\sigma_{\lambda 0} \cong 0.525$ nm and $\beta_{30} = 0.065$ ps$^3$/km. For comparison, the black bold line represents the gain spectrum without variations of $\lambda_0$, i.e. $\sigma_{\lambda 0} = 0$. Note that a gain reduction occurs at signal wavelengths at the center of the gain spectrum ($\Delta\omega_s = 0$) and at the outer peaks (where $\kappa = 0$); no gain variation occurs for signal wavelengths at the pumps. Interestingly, at $\Delta\omega_s \sim \pm\sqrt{-6\beta_{2c}/\beta_4}$ the gain increases slightly, resulting in a flatter spectrum. Note that since the pumps are optimized to obtain the flattest gain, their locations do not necessarily coincide with those that give the minimum ripple when $\sigma_{\lambda 0} = 0$.

For large values of $\sigma_{\lambda 0}$ we observed, as expected, a strong gain reduction at the center of the spectrum resulting in a useless FOPA [25-27].

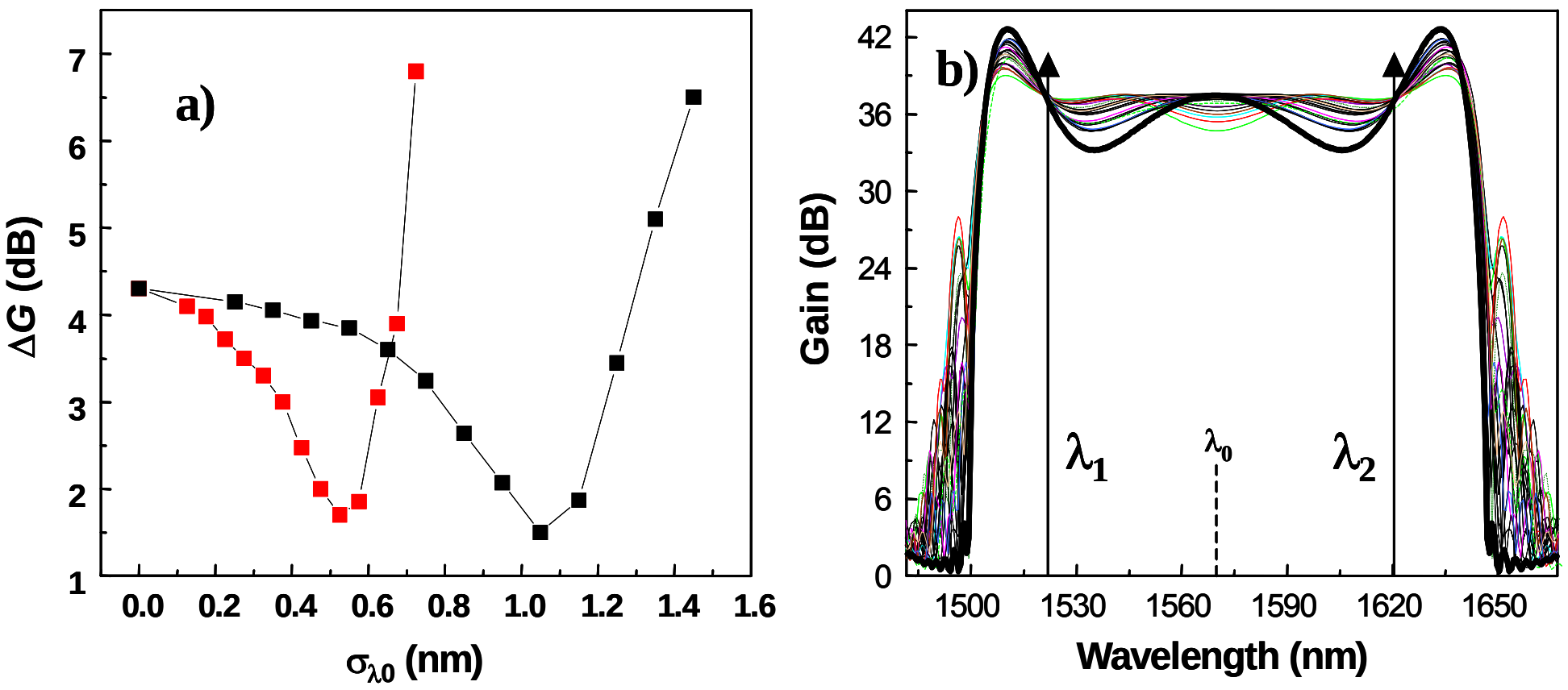

Fig. 6. (a) Gain ripple as a function of $\sigma_{\lambda 0}$ for $\beta_{30} = 0.065$ ps$^3$/km (red squares) and $\beta_{30} = 0.0325$ ps$^3$/km (black squares). The lines are guides for the eye. (b) Optimized output spectra obtained with 25 simulated fibers having $\sigma_{\lambda 0} = 0.52$ nm and $\beta_{30} = 0.065$ ps$^3$/km. The bold black line indicates the spectrum for the case $\sigma_{\lambda 0} = 0$.

### *5.2 The influence of $\gamma(P_1 + P_2)$, $L_{corr}$, and $\Delta\omega_p$ on the impact of $\lambda_0$ fluctuations*

We analyze now the influence of the correlation length by considering the same 2P-FOPA as in the previous subsection (i.e. $\gamma(P_1 + P_2) = 28$ km$^{-1}$, $L = 0.2$ km, $\beta_{30} = 0.065$ ps$^3$/km, $\beta_4 = -2 \times 10^{-4}$ ps$^4$/km, $\langle\lambda_0\rangle = 1570$ nm, and pumps separation ~100 nm), but now we change $L_{corr}$ to 8.65 m. Our results are plotted in Fig. 7 by the cyan triangles. Again, the gain ripple exhibits the same behavior: decreases as $\sigma_{\lambda 0}$ increases reaching a minimum value for $\sigma_{\lambda 0} = 0.71$ nm before increasing strongly. For comparison, we have plotted in red squares the case with $L_{corr}$ = 86.5 m. Note that decreasing $L_{corr}$ by a factor of 10 allows $\sigma_{\lambda 0}$ to increase by a factor of 1.4 in order to have the same impact on gain ripple. This result indicates that the dependency on $L_{corr}$ is much smaller than that with $\beta_{30}$.

Now we turn our attention to analyze the impact of pump separation. The 2P-FOPA parameters are: $\gamma(P_1 + P_2) = 28$ km$^{-1}$, $L = 0.2$ km, $\beta_{30} = 0.065$ ps$^3$/km, $\langle\lambda_0\rangle = 1570$ nm, $L_{corr} = 86.5$ m, $\beta_4 = -1.25 \times 10^{-5}$ ps$^4$/km, and pumps separation ~200 nm. The value of $\beta_4$ was reduced in order to keep constant $\xi$. The blue triangles in Fig. 7 show the results. Note that the minimum ripple is obtained for $\sigma_{\lambda 0} = 0.13$ nm. Comparing with the case of pumps separation of 100 nm (red squares), it is noted that an increase of the pump separation by a factor of two, in order to have the same impact of variations of $\lambda_0$ on the gain ripple, the fiber should have a value of $\sigma_{\lambda 0}$ four times smaller.

Finally, we change $\gamma(P_1 + P_2)$ to 56 km$^{-1}$ and the 2P-FOPA parameters are now: $L = 0.1$ km, $\beta_{30} = 0.065$ ps$^3$/km, $\langle\lambda_0\rangle = 1570$ nm, $L_{corr} = 86.5$ m, $\beta_4 = -2.5 \times 10^{-5}$ ps$^4$/km, and pumps separation ~200 nm. The value of $\beta_4$ was reduced in order to keep $\xi$ constant. The green circles in Fig. 7 show the results. Comparing with the case of $\gamma(P_1 + P_2) = 28$ km$^{-1}$ (blue triangles), it is noted that an increase of $\gamma$ by a factor of two, in order to have the same impact of variations of $\lambda_0$ on the gain ripple, the fiber should have a value of $\sigma_{\lambda 0}$ two times larger.

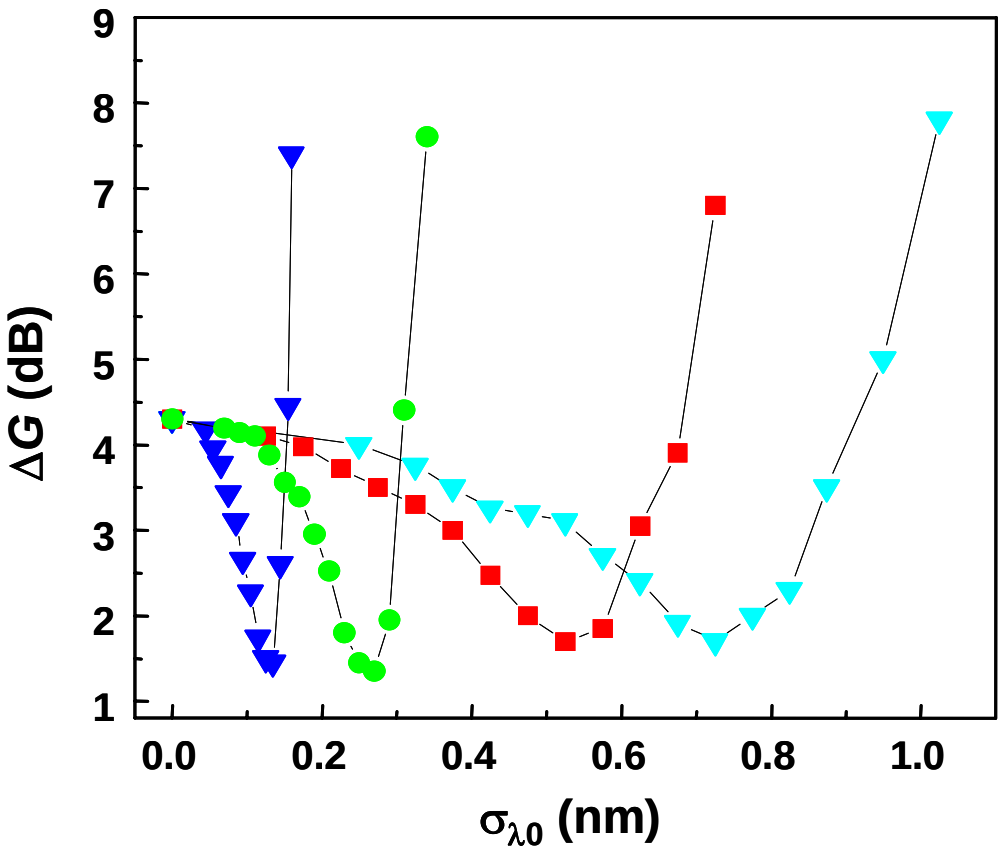

Fig. 7. Gain ripple as a function of $\sigma_{\lambda 0}$ for $\gamma(P_1 + P_2) = 28$ km$^{-1}$, $L = 0.2$ km, pumps separation ~100 nm, and $L_{corr} = 8.65$ m (cyan triangles). For $\gamma(P_1 + P_2) = 28$ km$^{-1}$, $L = 0.2$ km, pumps separation ~200 nm, and $L_{corr} = 86.5$ m (blue triangles). For $\gamma(P_1 + P_2) = 28$ km$^{-1}$, $L = 0.1$ km, pumps separation ~200 nm, and $L_{corr} = 86.5$ m (green triangles). The lines are guides for the eye. In all cases $\beta_{30} = 0.065$ ps$^3$/km and $\beta_4$ is varied in order to keep $\xi$ constant. For comparison the data with red squares in Fig. 6(a) is plotted.

The numerical simulations in Figures 6 and 7 showed the influence of the various parameters on the impact of $\lambda_0(z)$ in 2P-FOPA gain. Similar conclusions can be derived from taking the derivative of $G$ with respect to $\omega_0$. To have tractable expressions it is convenient to consider that in the region of high parametric gain, $\kappa / 2\gamma P_0 << 1$. In this limit $\log(1-\left(\frac{\kappa}{2\gamma P_0}\right)^2) \sim 0$ and $x \approx x_0\left(1-\kappa^2/8\gamma^2 P_0^2\right)$. Then the parametric gain can be written as $G_{\mathrm{dB}} \sim 8.7x - 6$. The gain fluctuation, $\delta G$, due to a variation of $\delta\omega_0$ in $\omega_0$ is then

$$\delta G \approx \frac{8.7 L \kappa \beta_{30}(\omega_s - \omega_1)(\omega_s - \omega_2)}{4\gamma P_0}\delta\omega_0 \tag{15}$$

The amount of gain variation is proportional to $\beta_{30}$, $\delta\omega_0$, and $L$ and inversely proportional to $\gamma P_0$. $\delta G$ depends also on the signal wavelength location: signal wavelengths close to the pumps suffer low gain variations, while signal wavelengths far from both pumps suffer of larger gain variations. This behavior is in agreement with results shown in Figs. 6 and 7. Eq. (15) also indicates that signal wavelengths where there is phase matching are less affected by variations of $\omega_0$. This is in disagreement with the findings in Fig. 6(b).

*5.3 The impact of polarization mode dispersion (PMD)*

PMD produces a misalignment of the states of polarization of the pumps and the signals changing the FWM efficiency as these waves propagate along the fiber. The alignment of pumps can be quantified by the internal product of their polarization vectors $\mathbf{s}(\omega)$. If they have parallel states of polarization at the fiber input then at the fiber output their internal product is given by $\langle s(\omega_1).s(\omega_2)\rangle = \exp\left[-4D_p^2 L\Delta\omega_p^2/3\right]$, where $D_p$ is the PMD coefficient, and $L$ the fiber length [33]. The depolarization effects can be related to a diffusion length defined as $L_d = 3/(2D_p\Delta\omega_p)^2$, where $L_d$ indicates the distance for which the scalar product is reduced from 1 to 0.37 (i.e. by 4.3 dB) [29]. It was shown through numerical simulations that, as a rule of thumb, if $L_d > L$, then the PMD decreases the gain spectrum uniformly without distortions [29]. If $L_d < L$, the PMD reduces the gain and induces distortions in the shape of the gain. These distortions are more pronounced at the center of the gain spectrum because the polarization of signal and pumps exhibit more misalignment. The impact of PMD, for a FOPA for which $L_d > L$, can be easily taken into account simply by considering in Eq. (2) an effective interaction length over which the polarizations of pumps and signals, are aligned. In section 7 we show experiments for which $L_d < L$.

## 6. Experimental setup and experimental results: short length fibers

We built 2P-FOPAs using three different fibers, A, B, and C, whose parameters are quoted in Table II. Fig. 8 shows the experimental setup. We used tunable external cavity lasers at $\lambda_1$, $\lambda_2$, and $\lambda_S$ as pumps and signal sources. In the case of fibers A and B, the pumps were amplified using C-band or L-band EDFAs. In order to obtain high power from the EDFAs, the pump lasers were amplitude modulated in the form of pulses with durations in the range 5-45-ns. We used an additional short length of fiber as relative delay ($\tau$) between the pump pulses to compensate for differences in optical paths, so that, within the FOPA fiber, the two pulses overlapped in time within 5 % of the width. Optical filters (OF) were used to reject most of the ASE from the EDFAs. Polarization controllers (PCs) were used to align the states of polarization of pumps and the signal so as to maximize the parametric gain. The spectra were characterized using an optical spectrum analyzer (OSA) with 0.1 nm resolution, and the peak pump powers were measured using a photodiode and a fast oscilloscope. The fibers were selected after estimating the value of $\sigma_{\lambda 0}$ with the method reported in [32]. We estimate the error in the gain measurements to be ±0.7 dB.

In the case of fiber C, pump 1 was obtained using a single pumped FOPA made with a HNLF having $L$ = 35 m and pumped with ~30 W pulses as indicated in Fig. 8 with the dotted lines. Using this approach we were able to obtain up to 4 W peak powers at these wavelengths - more than enough to pump the 2P-FOPA. To select this pump 1 we used a WDM coupler that filtered out wavelengths larger than 1515 nm. Figure 8(b) shows an example of a 2P-FOPA output spectrum measured in fiber C with $L$ = 150 m. Note that amplified noise around the pumps comes from the noise (that was unfiltered with the WDM) generated in the 1P-FOPA. ‘Spurious’ FWM tones that are 26 dB smaller than the signals can be also observed.

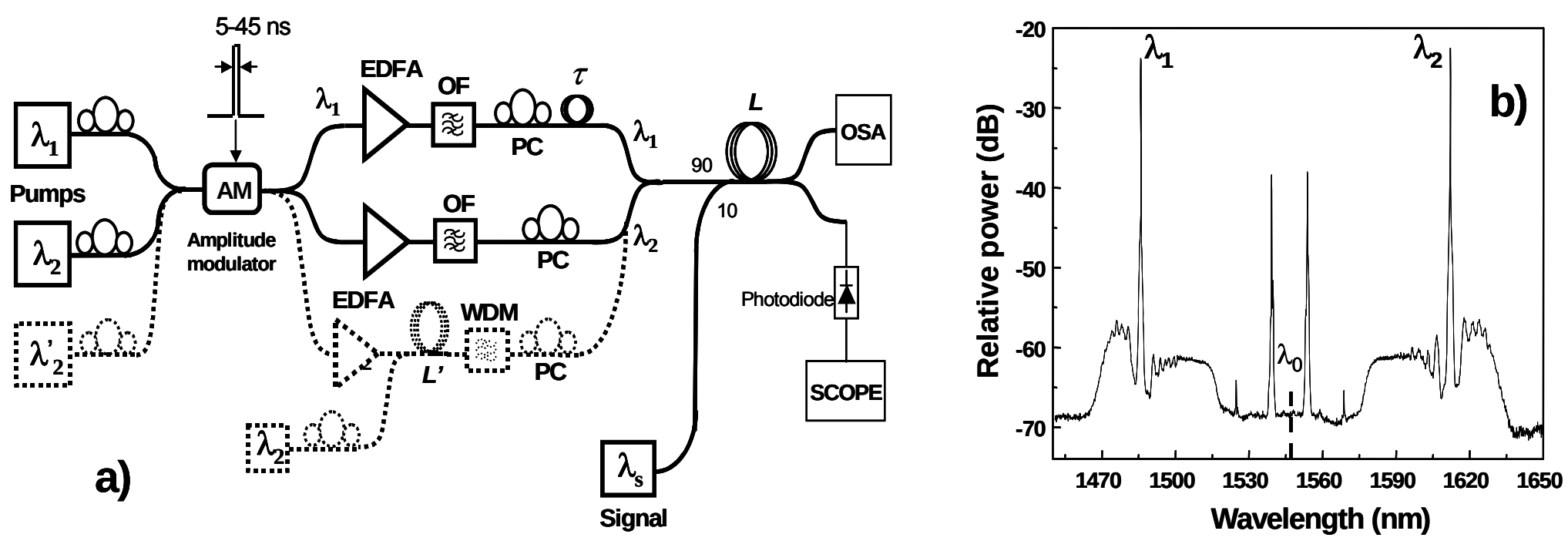

Fig. 8. (a) Experimental setup for measurements of gain in FOPAs. (b) 2P-FOPA output spectrum measured in fiber C with $L$ = 150 m.

Table II: Parameters for the three fibers in the experiments.

| Parameter | A | B | C |
|---|---|---|---|
| $L$ (m) | 950 | 300 | 150, 100 |
| $A_{eff}$ ($\mu m^2$) | 55 | 14.45 | 14.9 |
| $\gamma$ ($W^{-1}$/km) | 2.1 | 8.0 | 7.5 |
| $\alpha$ (dB/km) | 0.2 | 0.4 | 0.4 |
| $\beta_{30}$ ($ps^3$/km) | 0.12 | 0.065 | 0.073 |
| $\beta_4$ ($ps^4$/km) | $(-6 \pm 0.1)\times10^{-4}$ | $(-1.6 \pm 0.1)\times10^{-4}$ | $(-1.7 \pm 0.1)\times10^{-4}$ |
| $\langle\lambda_0\rangle$ (nm) | 1568.2 ± 0.05 | 1570.1 ± 0.05 | 1552.8 ± 0.05 |
| $\sigma_{\lambda 0}$ (nm) | ~0.1 | ~0.1 | ~0.1 |
| PMD (ps/√km) | ~0.03 | ~0.04 | ~0.04 |

*6.1 Conventional dispersion shifted fiber with $L_A$ = 0.95 km*

We measured the gain spectrum for three pump wavelength separations: 55, 62, and 68.9 nm. In order to keep the same gain in these three cases, the pump powers needed to be increased from $P_1 \cong P_2$ ~1.8 W (pump separation of 55 nm) to ~2.1 W (68.9 nm). The results are plotted with blue circles in Figs. 9 (a), (b), and (c), respectively. In each case the pumps locations were optimized to minimize the gain ripple. In this conventional DS fiber the spectral region for ripple minimization was 75-80 % of the region between the pumps. Note that the gain ripple increases as the pump separation increases from: $\Delta G \cong$ 3.3 dB for 55 nm pump separation to 5 dB for 68.9 nm. The diffusion lengths for each pump separation are $L_{d(a)}$ = 1.88 km, $L_{d(b)}$ = 1.48 km, and $L_{d(c)}$ = 1.2 km. In each case $L_d > L$, so we expect that the effect of any PMD would be to decrease the gain as the pump wavelength separation increases, but without introducing noticeable distortion in the gain spectra. We did simulations using Eq. 2 to compare with the experimental data. To take into account the possible effect of variation of $\lambda_0$ and PMD, we considered an effective interaction length $L_{int}$ that corresponds to the experimental gain for each pump separation. These lengths were: $L_{int}$ = 0.78 km, 0.73 km, and 0.67 km, respectively. The results are plotted in Figure 9 using black and red lines, for $\lambda_0$ = 1568.25 and 1568.15 nm, respectively. There is a very reasonable agreement between experiments and Eq. (2), meaning that real fibers, that are less than perfect, can be modeled with simple analytical expressions if longitudinal variations of $\lambda_0$ and PMD are sufficiently low.

Table III shows the values of $\Delta G_{dB}$ obtained using the simple expression derived by fitting $\Delta G$ (see caption in Fig. 3), together with the experimental values obtained by measuring $G_{max}$ and $G_{min}$ in a region ~ 75-80 % between the pumps.

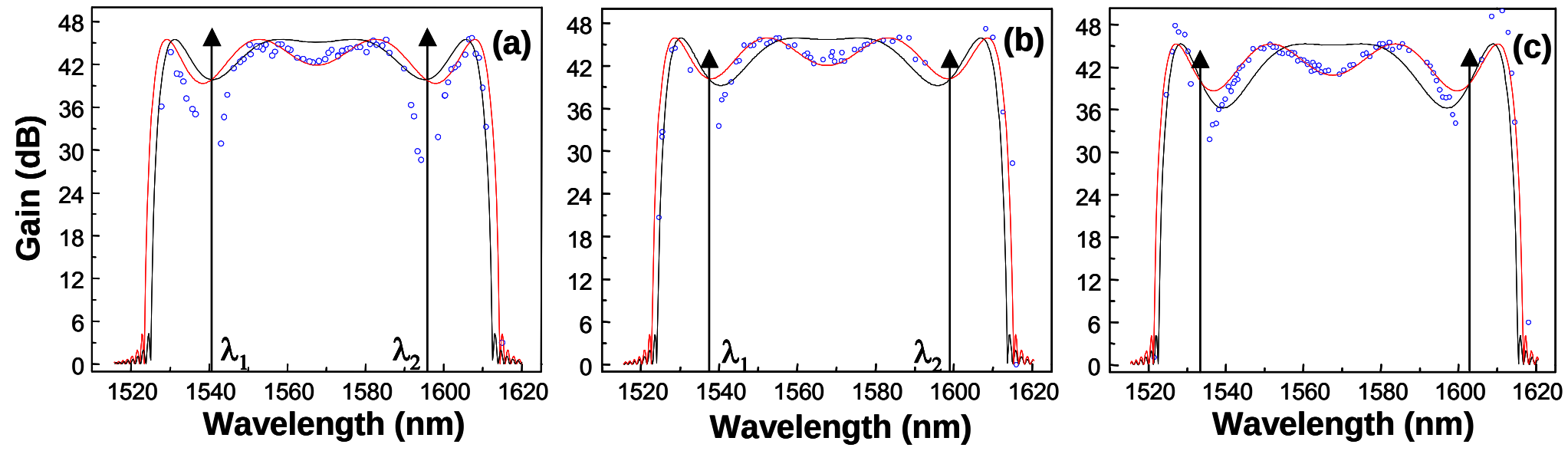

Fig. 9. 2P-FOPA gain spectrum for: (a) $\lambda_1$ = 1540.7 nm and $\lambda_2$ = 1595.65 nm; (b) $\lambda_1$ = 1537.3 nm and $\lambda_2$ = 1599.2 nm; (c) $\lambda_1$ = 1533.9 nm and $\lambda_2$ = 1602.8 nm. Blue circles: experimental points. Black and red lines: gain spectrum using Eq. 2 for $\lambda_0$ = 1568.15 nm and $\lambda_0$ = 1568.25 nm, respectively. The effective interaction lengths are (a) $L_{int}$ = 0.78 km. (b) $L_{int}$ = 0.73 km. (c) $L_{int}$ = 0.67 km.

Table III. Experimental and numerical $\Delta G$ for the three pump wavelength separations.

| Spectrum | $\lvert\xi\rvert$ | $\Delta G_{exp}$ (dB) | $\Delta G_{num}$ (dB) |
|---|---|---|---|
| a | 0.66 | 3.3 | 3.2 |
| b | 1 | 3.8 | 4.2 |
| c | 1.4 | 4.9 | 5.7 |

Two additional measurements were made to further characterize the 2P-FOPA. In the first, we verified that the measured gain was independent of which end of the fiber was used to input the signal. In the second measurement, we analyzed polarization dependent gain (PDG). The polarization states of pump1 and pump2 were adjusted to be perpendicular by minimizing the gain of ASE noise. The pump powers were set to $P_1 \cong P_2$ ~ 2.1 W, and the wavelength separation to 40 nm. The state of polarization of the signal was then varied in order to measure the maximum, $G_{max(pol)}$ and minimum gain $G_{min(pol)}$. The PDG = $G_{max(pol)} - G_{min(pol)}$ was measured in the spectral region between $\lambda_0$ and the pump at $\lambda_2$ (the region between $\lambda_0$ and $\lambda_1$ should be a replica of this due to symmetry). The PDG was around 2 dB that is low but not negligible. The same measurement was made for a pump separation of 69 nm, but we were unable to obtain a PDG smaller to 5 dB in the region between the pumps.

*6.2 Highly nonlinear dispersion shifted fiber with $L_B$ = 0.3 km*

The pumps were first located at $\lambda_1 \cong 1528.6$ nm and $\lambda_2 \cong 1613.75$ nm, while the pump powers were $P_1 \sim 1.9$ W and $P_2 \sim 1.3$ W. These values correspond to $\xi \cong -0.28$. The pump wavelengths were optimized to minimize the ripple in a spectrum having 5 extrema, as shown in Fig. 10(a) with blue circles. Note that high and flat gain ($G \cong 35 \pm 1.5$ dB) was obtained over 71 nm bandwidth. There is an appreciable tilt in the gain spectrum due to the Raman gain produced by the pump at $\lambda_1$ (the measured Raman gain at $\lambda_2$ is ~1.4 dB). Using the experimental parameters in Eq. (2) we obtained the gain spectrum for two values of $\lambda_0$: 1570.1 nm (red line) and 1570.15 nm (black line). The effective interaction length was 248 and 243 meters, respectively.

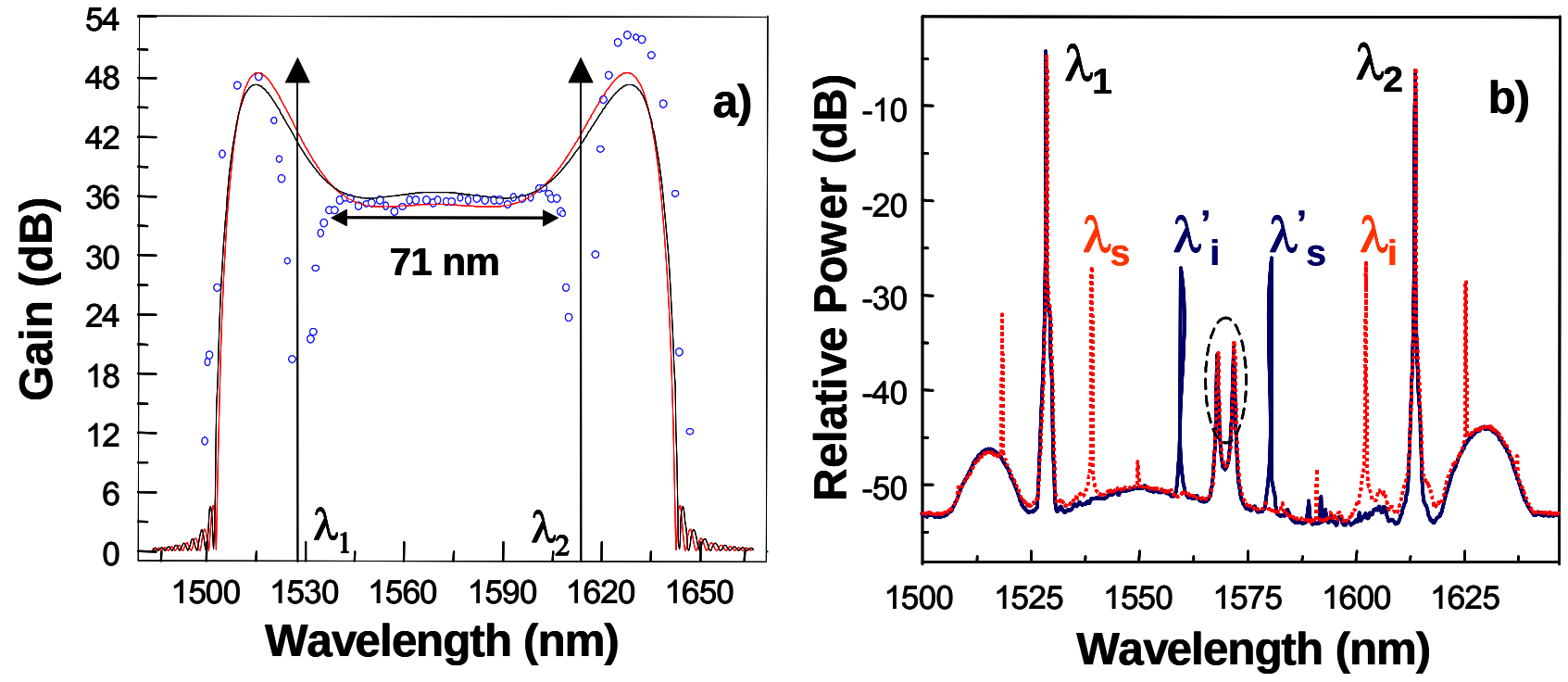

Fig. 10. (a) 2P-FOPA gain spectrum: blue squares (measurement), black line ($\lambda_0$ = 1570.1 nm), red line ($\lambda_0$: 1570.15 nm). (b) Output spectra for two locations of $\lambda_s$ = 1539 nm (red dotted line) and $\lambda'_s$ = 1581 nm (blue line). The ellipse indicates unfiltered noise due to the 40 nm free spectral range of the Fabry-Perot filter.

The measured and calculated ripple in a region $\Delta\omega_s = 0.83\Delta\omega_p$ (83 % of the region between the pumps) are $\Delta G_{exp}$ = 2.3 dB and $\Delta G_{calc}$ = 0.82 dB, respectively. The agreement is reasonable within the experimental error and the low impact of variations in $\lambda_0$ and PMD ($L_d$ = 0.44 km for this case). Fig. 10(b) shows typical output spectra for two signal locations: $\lambda_s$ = 1539 nm (red dotted line) and $\lambda'_s$ = 1581 nm (blue line).

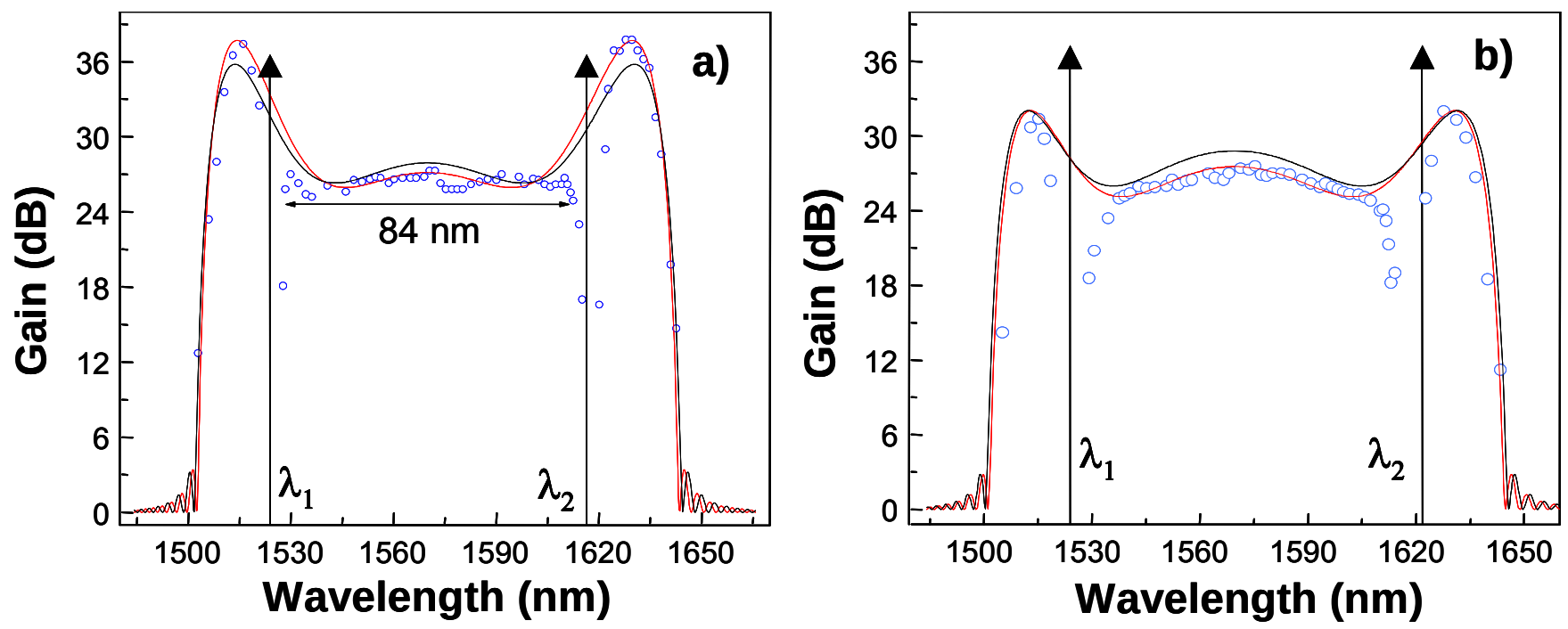

Fig. 11. 2P-FOPA gain spectrum. (a) Blue squares (measurement), black line ($\lambda_0$ = 1570.1 nm), red line ($\lambda_0$: 1570.05 nm). (b) Black circles (measurement), black line ($\lambda_0$ = 1570.15 nm), red line ($\lambda_0$: 1570.1 nm).

The pump separation was then increased to 93 nm in order to expand the bandwidth. In one experiment the pumps were located at $\lambda_1 \cong 1524.75$ nm and $\lambda_2 \cong 1617.75$ nm to minimize the gain ripple over the largest bandwidth. The pump powers were $P_1 \sim 1.7$ W and $P_2 \sim 1.1$ W. Figure 11(a) shows that $G \cong 26 \pm 1.5$ dB over 84 nm. We then used the experimental parameters to calculate the gain spectrum using Eq. 2. Two values of $\lambda_0$ were used to fit the data: 1570.05 nm (red line) and 1570.1 nm (black line). The effective interaction lengths were 230 and 220 meters, respectively, which should take into account the effects of PMD and longitudinal variations of $\lambda_0$. The agreement with experiments is quite good confirming that variations of $\lambda_0$ and PMD may decrease the gain, but without introducing distortions in the spectrum.

Note further in Fig. 11(a) that a consequence of having flat gain is reduction in the overall gain: the maximum gain (occurring at the outer signal wavelengths) is 10 dB higher compared to the parametric gain in the region between the pumps. Any attempt to increase the gain in the region between the pumps leads to an increased gain ripple. This is shown clearly in Fig. 11(b) where pump1 was detuned to $\lambda_1 = 1524.6$ nm and the pump power was decreased to have the same amount of gain (~26 dB). The 3 dB bandwidth decreased to 78 nm, whereas the difference between gain for outer and inner signal wavelengths decreased to 4.5 dB. Figure 11(b) also shows fittings to the experimental data for two values of $\lambda_0$: 1570.05 nm (black line) and 1570.1 nm (red line).

*6.3 Highly nonlinear dispersion shifted fiber with $L_C$ = 0.1 and 0.15 km*

To investigate the 2P-FOPA gain flatness in the case where the pumps are separated by more than 100 nm we used the fiber *C* (see Table II). This fiber had 2 km of length and was cut in several pieces, with lengths varying from 100 to 370 m and having estimated variations of $\lambda_0$ from ~0.1 to ~0.4 nm. Figure 12 shows gain spectra obtained with the fibers with the smallest variations of $\lambda_0$. The pump at $\lambda_1$ was generated using a 1P-FOPA. Figure 12(a) shows the gain spectrum of a fiber with 150 meters pumped with $P_1 \sim P_2 \sim 2.1$ W at $\lambda_1 \cong 1495.9$ nm and $\lambda_2 \cong 1611.9$ nm. We obtained high and flat gain, $G \cong 25 \pm 2$ dB, over ~102 nm. We also show two spectra calculated using Eq. (2) with $L_{\text{int}} = 119$ m and $\lambda_0 = 1552.73$ (black) and $\lambda_0 = 1552.78$ nm (red). With our parameters we have $\xi = -0.95$ and, from Eq. 11, we expect a ripple of $\Delta G$ = 2.4 dB.

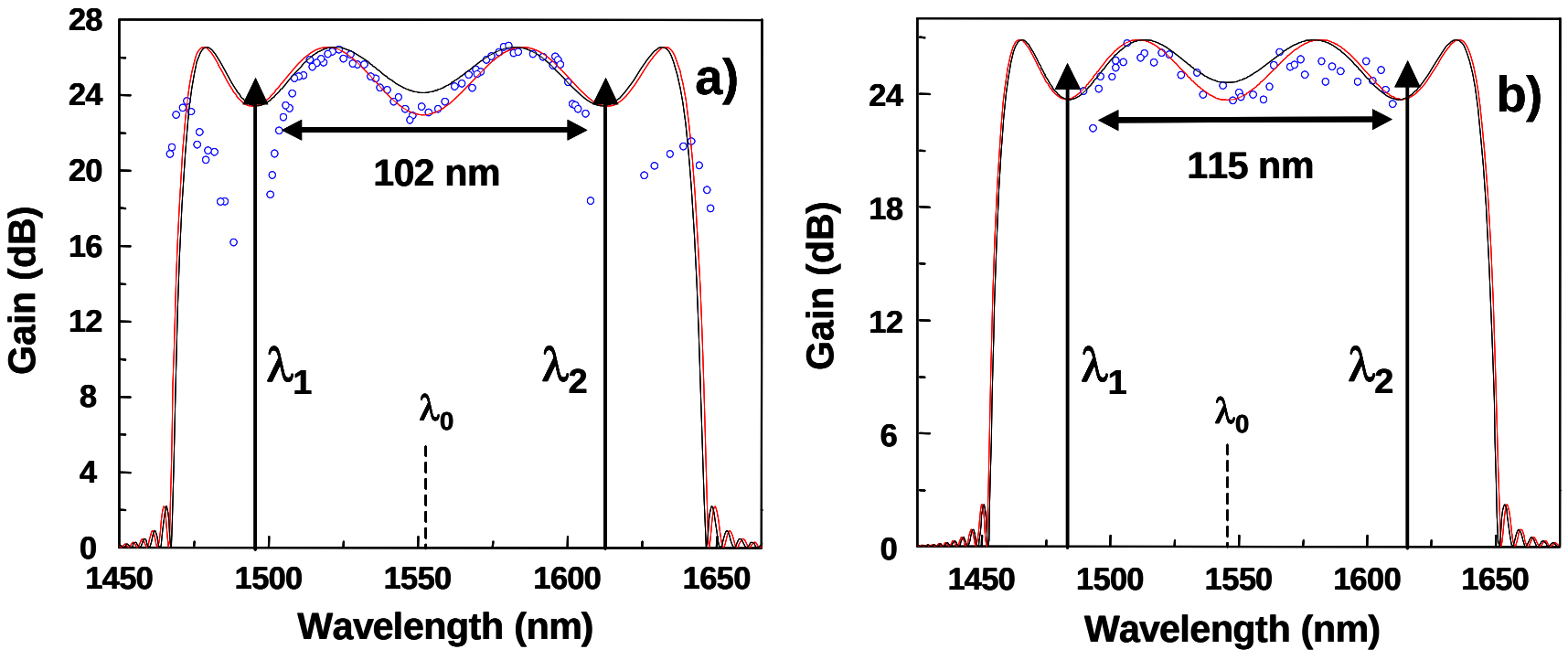

Fig. 12. 2P-FOPA gain spectrum measured with fiber C. (a) $L$ = 150 m. (b) $L$ = 100m.

The pump at $\lambda_1$ could be tuned over a large region because it was generated with a 1P-FOPA; however, the L-band EDFA limited the tunability of pump at $\lambda_2$ and as a consequence the 2P-FOPA bandwidth. To further increase this bandwidth we cooled the fiber with liquid nitrogen and the $\lambda_0$ was shifted to 1546.8 nm. Figure 12(b) shows the gain spectrum of a cooled fiber with $L$ = 100 m pumped with $P_1 \sim P_2 \sim$ 3.3 W. The pumps were at $\lambda_1 \cong$ 1483.1 nm and $\lambda_2 \cong$ 1613.6 nm and were again optimized in order to have the smallest gain ripple. Note that high and flat gain, $G \cong 25 \pm 1.5$ dB, over ~115 nm. This is, to the best of our knowledge, a record performance in terms of amount of gain and flatness for an optical amplifier. Dotted lines show fittings to the experimental data using Eq. (2) with $L_{\text{int}}$ = 76 m and using $\lambda_0$ = 1546.89 (red) and $\lambda_0$ = 1546.84 nm (black).

The good agreement between experiments and the simple analytical theory observed in Fig. 12 indicates, even for pump separations larger than 120 nm, the good quality of the HNLF in terms of low PMD and small fluctuations of $\lambda_0$. This was further confirmed in our experiments: by tuning slightly one of the pump we could retrieve the different spectral shapes (with 7 and 5 extremes) as in Fig. 1. Also, we have measured gain spectra for pump separations larger than 120 nm with the other fibers. Even for a variation of $\lambda_0$ of $\sigma_{\lambda 0}$ ~0.4 nm (fiber length of 370 m) we still observed gain spectra that were in good agreement with the theory.

## 7. Gain spectrum in long length fibers ($L_D$ = 13.8 km)

One motivation for using long fiber lengths is to use the 2P-FOPA as distributed amplifier and distributed wavelength converter. The parameters of this fiber are the same that fiber A, but now $\sigma_{\lambda 0}$ ~ 0.25 nm. The experimental setup is similar to that shown in Fig. 8; however, instead of using the amplitude modulator we used a phase modulator driven by three sinusoidal electrical signals (0.41, 1, and 2.4 GHz) in order to suppress the stimulated Brillouin scattering (SBS). We estimate the error in the measurements with this setup to be around ±0.5 dB.

The fiber was pumped with $P_1 \cong$ 190 mW and $P_2 \cong$ 170 mW and the gain spectrum was measured for three pump wavelength separations of $\lambda_2 - \lambda_1$ = 18.3 nm, 24.8 nm, and 39.4 nm. The pumps were also tuned in order to minimize the gain ripple. The results are shown in Figs. 14(a), (b), and (c), respectively. Note that as $\lambda_2 - \lambda_1$ increases, the ripple (calculated in the region between the pumps) increases and the amount of gain decreases strongly: $G = \langle G \rangle \pm \frac{1}{2}\Delta G$ = 36.5 ± 1.3 dB for $\lambda_2 - \lambda_1$ = 18.3 nm, 31 ± 1.5 dB for $\lambda_2 - \lambda_1$ = 24.8 nm, and 14.5 ± 4.5 dB for $\lambda_2 - \lambda_1$ = 39.4 nm.

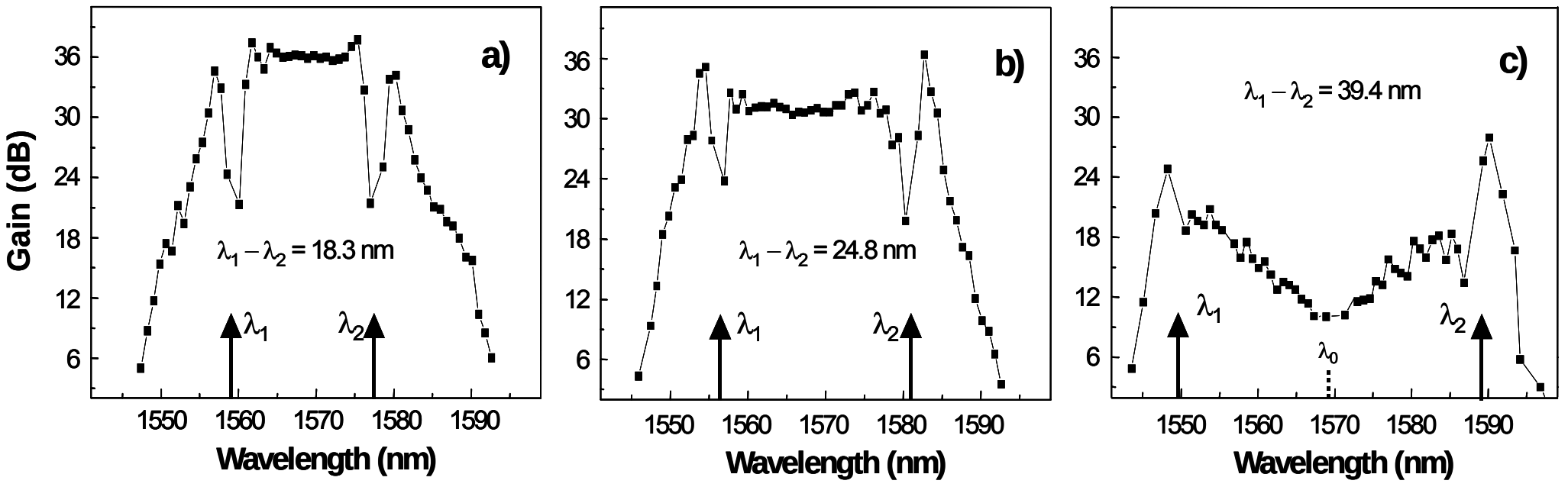

Fig. 14. 2P-FOPA gain spectra using fiber D for $P_1 \cong$ 190 mW and $P_2 \cong$ 170 mW (a) $\Delta\lambda_{\text{pumps}}$ = 18.2 nm, (b) $\Delta\lambda_{\text{pumps}}$ = 24.8 nm, and (c) $\Delta\lambda_{\text{pumps}}$ = 39.4 nm. The lines are guides for the eye.

If the gain is calculated using Eq. (2), we find that for the region between the pumps $G$ = 51 ± 0.3 dB for $\lambda_2 - \lambda_1$ = 18.3 nm and $G$ = 50 ± 3 dB for $\lambda_2 - \lambda_1$ = 39.4 nm. The disagreement between Eq. (2) and the experimental data is considerable and is likely related to both longitudinal variations of $\lambda_0$ and PMD. Since this fiber has a long length, it is reasonable to suppose that PMD will produce a considerable misalignment of pump and signal polarizations, thus reducing the gain. The calculated diffusion lengths for these pump separations are $L_d$ = 20.9 km, 11.4 km, and 4.5 km, respectively. These values of $L_\text{d}$ indicate that PMD could uniformly decrease the gain spectrum for the case $\lambda_2 - \lambda_1$ = 18.3 nm and also introduce increased distortions as $\lambda_2 - \lambda_1$ increases to 24.8 nm and 39.4 nm. To assess the contribution that variations of $\lambda_0$ have on the observed gain reduction, we performed numerical simulations including the estimated variations of $\lambda_0$. For the case $\lambda_2 - \lambda_1$ = 18.3 nm we found now that $\langle G \rangle$ drops to around 43-45 dB. This indicates that we can roughly attribute to variations of $\lambda_0$ as being responsible for 5-7 dB gain reduction. Numerical simulations were also performed for the case $\lambda_2 - \lambda_1$ = 24.8 nm and we found that $\langle G \rangle \sim$ 42-45 dB. Therefore, as the pump separation increases the main contribution for gain reduction is PMD.

## 8. Conclusions

We have studied numerically and experimentally broadband double-pumped fiber optical parametrical amplifiers (2P-FOPAs) having flat spectral response. Expressions for the gain ripple as a function of the FOPA parameters were deduced for the most representative kinds of 2P-FOPA spectra, which classified by their number of extrema. The impact that longitudinal variations of the zero dispersion wavelength has on the gain spectrum was studied in detail through numerical simulations. We showed that adequate amounts of variations of $\lambda_0$ tend to flatten the gain spectrum. We show that this amount of variation depends on $1/\beta_{30}$, on $\gamma(P_1 + P_2)$, on $L_\text{corr}$, and on $1/\Delta\omega_p^2$. We experimentally showed that by using well-designed highly non-linear fibers, 2P-FOPAs with flat spectral response over 115 nm can be obtained. Further improvement of fibers in terms of the value (and sign) of the fourth order dispersion and nonlinear coefficients would lead to 2P-FOPAs with flat operation over several hundreds of nanometers and without requiring pump powers larger than 0.5 W.

We stress that 2P-FOPA for real applications should use cw pumps and should also be implemented in order to have low PDG. Our measurements with co-polarized pulsed pumps, however, exemplify well the potentialities of this device in terms of bandwidth.

## Acknowledgments

We thank fruitful discussions with G.S. Wiederhecker. We thank Prof. Hypolito J. Kalinowski from CEFET-Paraná for making the Bragg gratings used in some of the experiments. We gratefully acknowledge J.B. Rosolem, A.A. Juriollo, C. Floridia, A. Paradisi, F. Simoes, and R. Arradi from CPqD Foundation for the loan of equipment used in this investigation. This work was financially supported by the Brazilian agencies Fapesp, Capes, and CNPq.

## Appendix A

We analyze the amplification of 80 WDM signal channels located from 1478 nm to 1538 nm (with 100 GHz spacing and –30 dBm input power) by solving the non-linear Schrödinger equation (in order to take into account all FWM processes). The parameters are: fiber length $L$ = 60 meters, $\lambda_0$ = 1543 nm, $|\lambda_1 - \lambda_2| \cong 163.2$ nm, $\gamma(P_1 + P_2) = 52.5$ km$^{-1}$, $\beta_{30} = 0.016$ ps$^3$/km, and $\beta_4 = -5.2\times10^{-5}$ ps$^4$/km. With these values, $x_0 = 3.15$ and $\xi = -0.72$. Figure A1 shows the spectra at the input and at the output of the 2P-FOPA: the WDM signals experience 20 dB of gain with a ripple of ± 1.4 dB over the 60 nm of bandwidth. The flattest gain spectrum is obtained when the pumps are located at $\lambda_1$ = 1462.37 nm and at $\lambda_2$ = 1625.55 nm. The red line in figure A1 shows the gain calculated using the Eq. 2 with identical parameters as in the NLSE. One important feature in Figure A1 is that spurious tones around the pumps and at the outer maxima are efficiently generated and are strong enough to produce considerable cross-talk [35]. In fact, in the NLSE simulation, the channels were located from 1478 nm up to 1540 nm, because for wavelengths smaller than 1478 nm the tones due to pump-signal spurious FWM are only ~15 dB smaller than signal and would introduce prohibited distortion. In the NLSE simulation we assumed pump and signal with parallel polarizations, which would normally enhance the generation of spurious tones; however we also used a low output signal power (~ –10 dBm).

The extent of the region of high crosstalk will depend on $\beta_3$, $\Delta\omega_p$, and the number of WDM channels. In our example, this forbidden band is ~20 % of the bandwidth between the pumps.

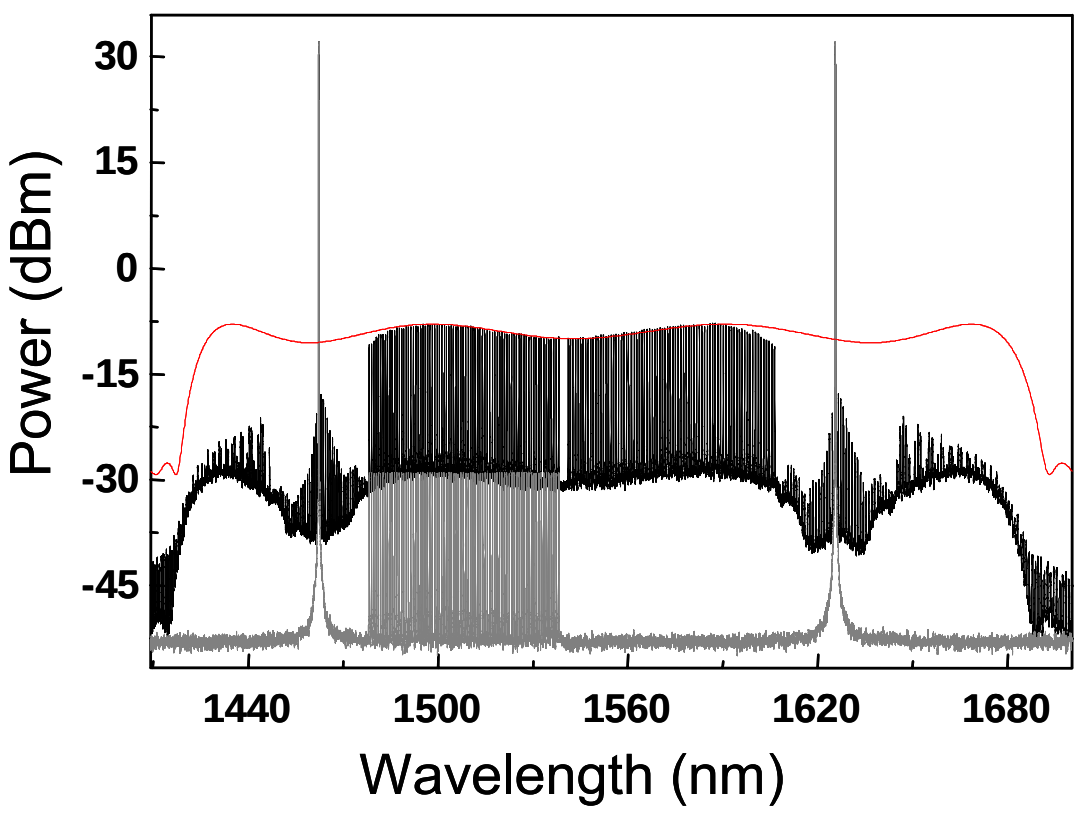

Fig. A1. Solid lines: Numerical solution of the NLSE for the amplification of 80 signals by a 2P-FOPA. Red line: Gain spectrum obtained with Eq. (2) using identical parameters in the NLSE.

## Appendix B

### *B.1 Calculation of gain ripple in spectrum having 5 extrema and $\beta_4$ > 0*

Spectra having 5 extrema in fibers with $\beta_4 > 0$ are differentiated from fibers with $\beta_4 < 0$ (which were analyzed in section 4), by the fact that the four roots of $\kappa = 0$ can be located in the region between the pumps. Thus, the maximum gain of this spectrum is now $G_{max} = 8.7x_0 - 6$. To minimize the gain ripple we need to maximize the gain at $\Delta\omega_s = \pm\sqrt{-6\beta_2/\beta_4}$ (local minimum). This implies in maximizing the gain at $\Delta\omega_s = 0$; i.e. setting $\kappa(\Delta\omega_s = 0) = 0$, from which we obtain

$$\beta_{2c} = \Delta\omega_p^2\beta_4(1-2\xi)/(24\xi)\,. \tag{B1}$$

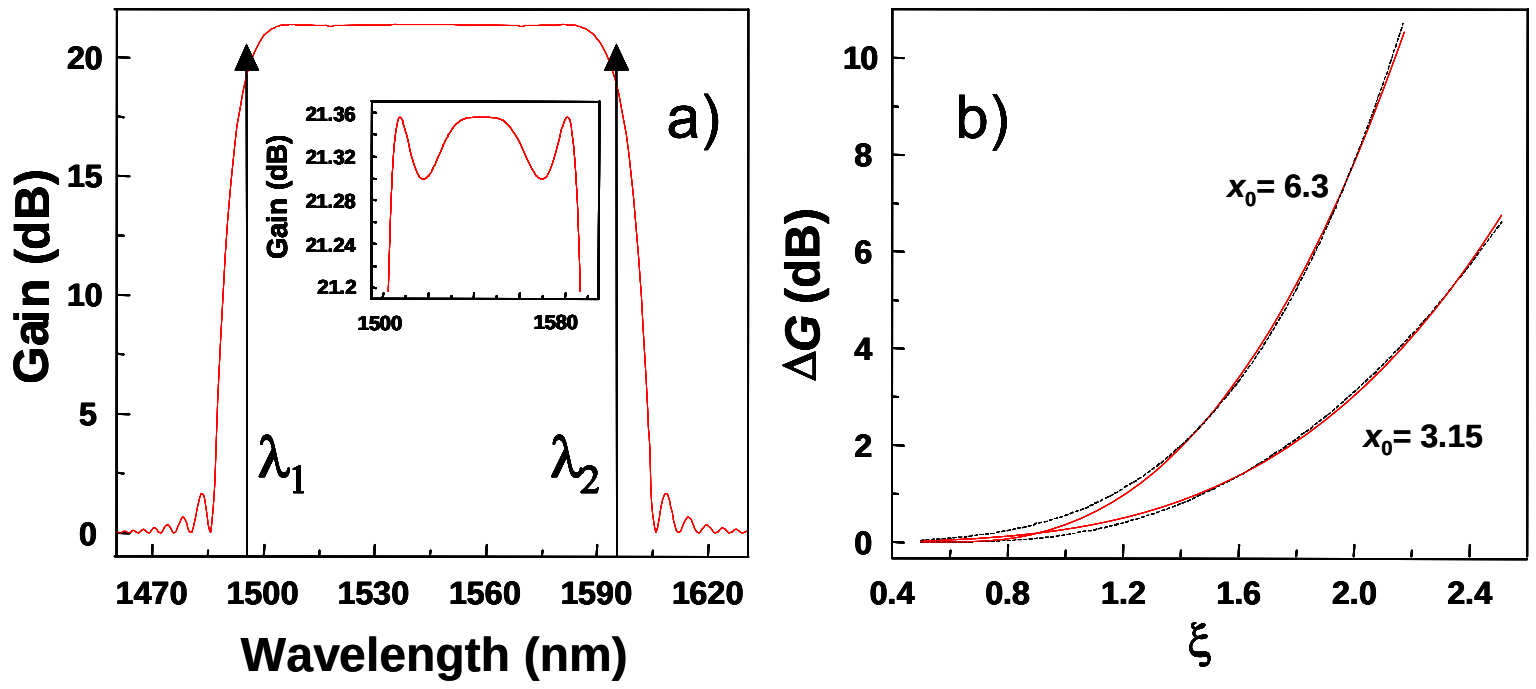

Fig. B1. (a) Gain spectrum obtained with $\beta_{2c}$ from Eq. B1. Inset: zoom of gain spectrum. (b) Gain ripple as a function of $\xi$ for two values of $x_0$. Continuous lines in red: analytical calculation. Dashed lines: power law fits to $\Delta G$. We have $\Delta G = 0.54\xi^{3.85}$ for $x_0 = 3.15$ and $\Delta G = 0.26\xi^{3.55}$ for $x_0 = 6.3$.

Figure B1(a) shows the gain spectrum obtained with $\beta_{2c}$ in Eq. (B1) with identical parameters as in Figs. 2-4 and for $x_0 = 3.15$. With this value of $\beta_{2c}$ it is easy to calculate the gain at $\Delta\omega_s = \pm\sqrt{-6\beta_{2c}/\beta_4}$ , and then $\Delta G$. The result is shown in Fig. B1(b).

*B.2 Calculation of gain ripple in spectrum having 1 extremum and $\beta_4 > 0$*

By looking at Eq. (4) it can be deduced that spectra having 1 extremum only occurs when both $\beta_{2c}$ and $\beta_4$ are positive. The condition to calculate $\Delta G$ in this spectrum is: 1) maximizing the gain at $\Delta\omega_s = 0$, then $G_{max} = = 8.7x_0 - 6$ and 2) calculate the gain at a frequency $\Delta\omega_s = b\Delta\omega_p$. The value of the phase mismatch at $\Delta\omega_s = b\Delta\omega_p$ is $\kappa_{min}/2\gamma P = b^2[0.5 + \xi(b^2 - 1)]^2$. Replacing this in Eq. (7) leads to the calculation of $G_{min}$. Figure B2 shows the plot of $\Delta G$ as a function of $\xi$. Note that spectra having only one extremum implies in $\xi$ restraint to $0 < \xi < 0.5$.

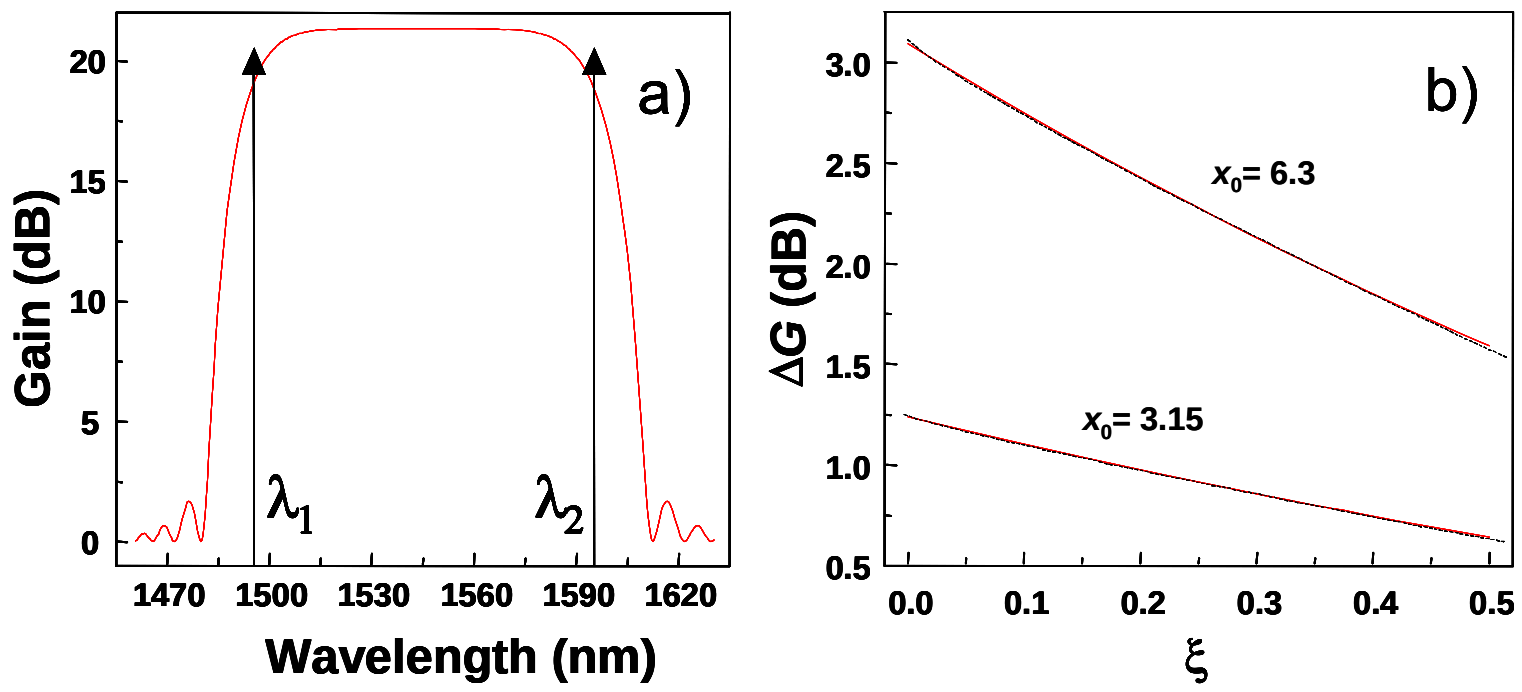

Fig. B2. (a) Spectrum with one extreme ($\beta_4 > 0$). (b) $\Delta G$ as a function of $\xi$ for two values of $x_0$. Continuous lines in red: analytical calculation. Dashed lines: power law fits to $\Delta G$. We have $\Delta G = 3.1 - 2.8\xi^{0.9}$ for $x_0 = 3.15$ and $\Delta G = 1.25 - 1.1\xi^{0.9}$ for $x_0 = 6.3$.